\documentclass[]{aastex631}

\usepackage{multirow}

\usepackage{hyperref}
\hypersetup{
    colorlinks=true,
    linkcolor=blue,
    filecolor=magenta,      
    urlcolor=cyan,
    pdftitle={Overleaf Example},
    pdfpagemode=FullScreen,
    }

\shorttitle{A High-resolution, Inversion-Based Synoptic Study of Solar Granulation}
\shortauthors{Crowley et al.}
\graphicspath{{./}{figures/}}
\begin{document}

\title{A High-resolution, Inversion-Based Synoptic Study of Solar Granulation}

\author[0009-0004-3717-6541]{James Crowley}
\altaffiliation{George Ellery Hale Graduate Fellow, CU Boulder}
\affiliation{National Solar Observatory, 3665 Discovery Dr, Boulder, CO 80303, USA}
\affiliation{University of Colorado, Boulder, Boulder, CO 80309}

\author[0000-0002-0189-5550]{Ivan Milic}
\affiliation{Institute for Solar Physics (KIS), Georges-Kh\"{o}ler-Allee 401a, 79110 Freiburg im Breisgau, Germany}
\affiliation{Faculty of Mathematics, University of Belgrade,Studentski Trg 16, 11 000, Belgrade, Serbia}
\affiliation{Astronomical Observatory, Volgina 7, 11 060, Belgrade, Serbia}

\author[0000-0002-6116-7301]{Gianna Cauzzi}
\affiliation{National Solar Observatory, 3665 Discovery Dr,Boulder, CO 80303, USA}
\affiliation{INAF - Osservatorio Astrofisico di Arcetri, Firenze, Italy 50125}

\author[0000-0001-8016-0001]{Kevin Reardon}
\affiliation{National Solar Observatory, 3665 Discovery Dr,Boulder, CO 80303, USA}
\affiliation{University of Colorado, Boulder, Boulder, CO 80309}

\begin{abstract}

The convectively driven, weakly magnetized regions of the solar photosphere dominate the Sun's surface at any given time, but the temporal variations of these quiet regions of the photosphere throughout the solar cycle are still not well known. To look for cycle-dependent changes in the convective properties of quiet Sun photosphere, we use high spatial and spectral resolution spectropolarimetric observations obtained by the Hinode Solar Optical Telescope (SOT) and apply the Spectropolarimetric Inversions Based on Response Functions (SIR) code to infer physical conditions in the lower solar photosphere. Using a homogeneous set of 49 datasets, all taken at disk center, we analyze the temperature stratification and the line-of-sight velocities of the granules and intergranules over a period of 15 years. We use a k-means clustering technique applied to the spectral profiles to segment the granules and intergranules based on both intensity and velocity. We also examine the profile bisectors of these different structures and compare these to past analyses. Our results show fairly constant properties over this period with no clear dependence on the solar cycle. We do, however, find a slight increase in the photospheric temperature gradient during the declining phase of the solar cycle. 
Our findings could have significant implications for understanding the coupling between the quiet Sun atmosphere and the global solar dynamo.

\end{abstract}


\section{Introduction} \label{sec:intro}

The cyclic solar dynamo gives rise to measurable variations of many solar properties. Over an 11-year period, the large-scale structure of the Sun's global magnetic field reverses polarity, and the number and strength of magnetic features such as faculae, sunspots, and active regions wax and wane dramatically \citep{2015LRSP...12....4H}. The total solar irradiance also varies in phase with the 11-year magnetic cycle, with an excursion amounting to about 0.1\% \citep{2016JSWSC...6A..30K}.  Models show that most of these changes are due to the different surface brightness of such magnetic features \citep{2017PhRvL.119i1102Y}; however, the underlying hypothesis that the thermodynamical state of the quiet-Sun atmosphere remains invariant through the cycle has not been fully proven to date. Indeed, the presence of a strong, cycling magnetic field in the solar interior is thought to modify the balance of convective and turbulent processes, leading to intrinsic variations of fundamental parameters such as the solar radius \citep{2007ApJ...670.1420M,2018AN....339..545S}, or the solar effective temperature, for which subtle variations correlated with the solar cycle have been reported \citep{1988Sci...242..908K,1997ApJ...474..802G}. Numerous authors have also described solar cycle-related variations of global helioseismic parameters \citep[cf. the review by][]{2016LRSP...13....2B},
which have led to the conclusion that the subsurface structure of the Sun might change in phase with the magnetic activity \citep{2005ApJ...633L.149L,2020ApJ...903L..29W}. 

Given that properties of stellar convection are quite sensitive to changes in stellar parameters like the ones mentioned above \citep{2013A&A...557A..26M,2013ApJ...769...18T}, it appears plausible that cycle-related changes might be directly observed in the convection pattern at the solar surface. To this end, numerous authors have investigated the cycle dependence of supergranules, a pattern of surface convection appearing at spatial scales of order 10–50 Mm  \citep{1962ApJ...135..474L}. While there appears to be a decrease of the supergranular cell sizes in phase with activity maxima \citep{2004ApJ...616.1242D,2008A&A...488.1109M}, some discordant results have been provided over the years \citep[e.g.][]{2017ApJ...841...70C,2017A&A...598A..99R}. It bears remarking that supergranules show a very low intensity contrast, making the convective pattern detectable mostly via related signatures such as the distribution of horizontal flows within the cells or the concentration of magnetic flux at cell boundaries \citep[see e.g. the discussion in][]{2018LRSP...15....6R}. On the contrary, the pattern of surface granulation at smaller spatial and temporal scales (of order 1-2 Mm and several minutes, respectively), is quite easily identifiable via the large intensity contrast between upflowing hot granules and the colder, downflowing intergranular lanes \citep{2009LRSP....6....2N}. 

Early studies on the possible cycle dependence of granular properties have been conducted mostly using broadband images obtained at the Pic du Midi Observatory and seemed to point towards a measurable decrease of the granular size in antiphase with solar activity \citep{1983CRASB.296..265M,1986SoPh..107...11R,1988AdSpR...8g.159M}. These results, however, were not confirmed by a subsequent analysis on an extended series of the same data \citep{2007A&A...475..717M}; this was probably due to the insufficient photometric precision as well as varying noise, including that induced by atmospheric seeing, as well as instrumental characteristics \citep[see also the introduction in][]{2018A&A...616A..87M}. 
% removed this indeed, this one didn't flow for me when I was reading it...
%Indeed, a
As highlighted by \citet{1998ASPC..140..455R}, a necessary pre-requisite for reliable results on this topic is the availability of a long series of uniform, high-resolution data, where instrumental, calibration and noise effects are both minimized and accurately known.

Space-based instruments such as the Broadband Filter Imager (BFI) and the SpectroPolarimeter \citep[SP,][]{2013SoPh..283..579L} on the Solar Optical Telescope \citep[SOT,][]{2008SoPh..249..167T} onboard the Hinode spacecraft, or the Helioseismic Magnetic Imager \citep[HMI,][]{2012SoPh..275..207S} onboard the Solar Dynamic Observatory, now offer a unique opportunity to revisit this topic with long series of data that approach the ideal conditions outlined above. In the absence of noise from the Earth's atmosphere, these instruments routinely achieve high spatial resolution, well suited to studies of granulation (the resolution is consistently $\sim 0.32"$ for SOT instruments, and $\sim$ 1.0" for HMI, with 1" corresponding to 720 km at solar disk center). Equally important, these instruments have been obtaining well-calibrated, standard observations for over a solar cycle, allowing long-term investigation of the possible variation of solar properties. 
Using the blue continuum images obtained by SOT/BFI over almost a full decade (November 2006 - February 2016), \citet{2018A&A...616A..87M} investigated the possible variation of granular properties during the solar cycle, but could not identify any significant change in spatial scale or intensity contrast within the instrumental noise. They estimate that the minimum variation detectable in their data would be of the order of 3\%, both in intensity contrast and spatial scale, thus providing an upper limit to the possible cycle-related changes. By using a 10 years-long series of HMI continuum images, corrected for a variety of instrumental effects, \citet{2021A&A...652A.103B} found a variation of the average granular size of the order of 2\% during the solar cycle, with the smallest sizes occurring with about one year delay with respect to the maximum of activity. Their complementary analysis of continuum images obtained with the BFI/SOT on Hinode, partially overlapping in time with the HMI series, tentatively confirmed the result.

The spectral and polarization dimensions of the Hinode/SP data provide additional information and constraints for photospheric studies. \citet{2016A&A...595A..71F} were the first to use the data from the
Hinode synoptic irradiance program (HOP 79 and HOP 412, see Sect. \ref{subsec:data}) to investigate whether the gradient of the photospheric temperature bears any relationship with the solar cycle. 
Since both the temperature variation and local gradients are important measures of energy transport in the photosphere \citep{2014ApJ...788..151C}, they can provide insight on the possible changes of the thermodynamical properties of the quiet-Sun atmosphere due to the global magnetic cycle.

\citet{2016A&A...595A..71F} compared intensity images formed at constant continuum optical depth and derived their actual geometrical height of formation to obtain the temperature gradient between 0 and 60 km above the 500 nm continuum formation height ($\log\,\tau_{500}=0$). For very quiet, internetwork regions, they found a steeper temperature gradient around the time of solar maximum compared to solar minimum but curiously, this trend was observed only for the Northern hemisphere of the Sun, while no difference was uncovered for the Southern hemisphere. However, since these authors analyzed only two instances (days) of the synoptic program, a broader investigation of the full Hinode irradiance data set might be able to confirm or negate these results on a statistically more significant basis.

In this paper, we study a similar question by using a large subset of the synoptic irradiance data acquired by Hinode/SP over the last 15 years and employing spectropolarimetric inversions to obtain the distributions and stratifications of thermodynamic parameters in the lower photosphere. We focus on disk-center regions without strong magnetic features and study the variation of the inferred temperature gradient and line-of-sight velocity with the solar cycle.
In Sect. \ref{sec:methods}, we describe the data, and our procedures for applying the inversion code to non-magnetic regions of Hinode SOT/SP data and separating the data into granular cells and intergranular lanes. In Sect. \ref{sec:results}, we analyze the changes in thermodynamic parameters, and finally, in Sect. \ref{sec:discussion}, we present our conclusions. 

\section{Data and Methods} \label{sec:methods}

\subsection{Hinode data selection}\label{subsec:data}

The spectropolarimeter \citep[SP, ][]{2013SoPh..283..579L} mounted on the Solar Optical Telescope  \citep[SOT][]{2008SoPh..249..167T} aboard the Hinode satellite has provided high-resolution observations of the solar photosphere since its first light in 2006. It was designed to accurately infer thermodynamic and magnetic properties by observing the Fe I doublet at 630.15 and 630.25\,nm,
a pair of neutral iron lines with similar heights of formation but
different magnetic sensitivities. With a diffraction limit of $\approx\,0.3"$  ($\approx\,$ 200\,km on the solar surface) and spectral resolution of around $2\times10^5$, the SP allows studies of photospheric features with a powerful combination of spatial and spectral resolution, as well as extended temporal coverage. 
While Hinode SOT/SP offers a much smaller field of view (FoV) than full disk telescopes like SDO/HMI, it has a significantly higher spatial resolution, allowing better identification of solar features (granules, intergranular lanes, magnetic regions), and significantly better spectral resolution and spectral sampling, providing stronger constraints on the variation of the atmospheric properties with depth.  

For this project, we chose datasets from the two Hinode observing plans (HOPs) designed for synoptic studies of the quiet sun: HOP 79 and HOP 412. These plans observe monthly along the solar meridian with the same exposure time, slit scanning step size, and other observing properties to study variations in solar irradiance during different phases of the solar cycle. Data from these same HOPs have been used by \cite{buehler_quiet_2013}, \cite{2016A&A...595A..71F}, and \cite{2021A&A...652A.103B} to study the quiet-Sun photosphere in detail. For this work, we only chose datasets at the disk center to avoid any projection effects and possible hemispheric differences \citep[e.g.][]{2013EGUGA..1511672F, 2016A&A...595A..71F}. We verified the observed fields were near the disk center by cross-correlating them with co-temporal HMI observations. Since our study focuses only on thermodynamical variations in the quiet photosphere, we selected datasets away from any active regions or areas of strong magnetic fields.
% that did not contain strongly magnetized regions. 

Doing so, we obtained 49 homogeneous, spectropolarimetric scans that contain observations of the quiet Sun close to the disk center, spanning almost 15 years, from January 2008 to December 2022. All datasets have been acquired with an exposure time of 4.8 seconds, resulting in a polarimetric precision of approximately $2\times 10^{-3}$, and have pixel scales of $0.148"$ per pixel in the slit scanning direction and $0.316"$ per pixel along the slit. These pixel scales have been verified by recent cross correlation with HMI \citep[as in][]{fouhey_large-scale_2023}. The spectral sampling is 2.15 pm (21.5 m\AA)/pixel. From each HOP raster, we selected a subset of 165 slit steps for further analysis, for a total FoV of 51.8" x 24.5". The list of the observations is provided in 
Table \ref{table:datasets}.

\subsection{Contrast variations} \label{subsec:contrast_changes} %through solar cycle

\begin{figure}
    \centering
    \includegraphics[width=150mm,scale=1]{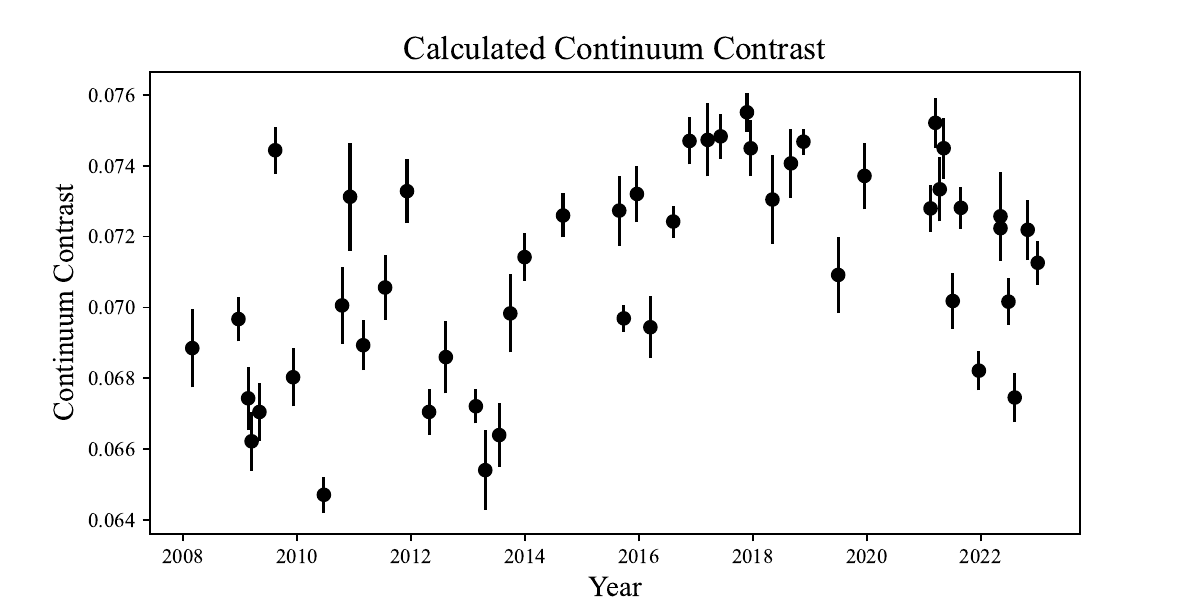}
    \caption{The calculated continuum contrast of each of the selected 49 datasets, showing the non-monotonic behavior of increasing and then decreasing over the 15 year interval. Error bars are calculated as described in \ref{subsec:temp_diff}.}
    \label{fig:contrast}
\end{figure}

Despite being chosen from the same observing program and having very similar heliocentric angles, the continuum contrast inferred from the SOT/SP data for the chosen maps varies between 6.4\% and 7.5\% over the 15-year range of the data (see Fig.\,\ref{fig:contrast}). 
%These measurements are an extension of similar contrast changes already seen by \citet{buehler_quiet_2013} and \citet{2014PASJ...66S...4L}. 
% commented out Jan 22- ask Kevin what this means: 
%While the variations have a similar magnitude, our independent dataset selection criteria result in different values and temporal trends in the contrasts. 
The cause of these changes is most likely the variation in the accuracy of the focusing of the SP \citep[see, e.g.][]{2014PASJ...66S...4L, hinode_review_team_achievements_2019}. \citet{Danilovic2008hinodevsmhd} have shown that a defocus of $\sim$1.5 mm (or 9 focusing steps) is sufficient to reduce the contrast of the quiet Sun measured by SOT/SP by 1\%, a range comparable to what we see in our datasets.
%needs to be accounted for to reconcile Hinode SOT/SP observations and synthetic spectra from MHD simulations. 
We found that these changes in focus have detectable effects on thermodynamic quantities derived from the data by mixing information from different types of photospheric structures and should be carefully considered in long-term studies using Hinode SOT data. To avoid confusing physical variations of the solar atmosphere with instrumental effects, we then chose to slightly blur all datasets so that the continuum contrast is equal between all 49 datasets, similar to what \citet{2014PASJ...66S...4L} and \citet{2015ApJ...807...70J} proposed. This additional blur is implemented as a spatially symmetric 2D Gaussian filter, taking into account the different spatial sampling in the two axes of the images. The width of the Gaussian is calculated for each dataset to achieve a reduced contrast equal to the lowest-contrast dataset (6.4\%). The full-width-half-maxima of the Gaussians needed to equalize the contrast range from 0" to 0.17", with an average of 0.14". We cannot completely rule out that the changes in the continuum contrast are, indeed, due to actual varying spatial structuring of the solar atmosphere, but given the good understanding of the variations in instrumental performance, we find it more reasonable to interpret these contrast variations as changes in focus.

\subsection{Spectropolarimetric Inversions} \label{subsec:inversions}

To derive the stratifications of thermodynamic parameters in the lower photosphere, we employed the well-known spectropolarimetric inversion code SIR  \citep[Stokes Inversions Based on Response Functions,][]{ruiz_cobo_inversion_1992}, which efficiently infers optical-depth-stratified thermodynamic and magnetic properties from each observed polarized spectrum, under the assumption of local thermodynamic equilibrium (LTE) and hydrostatic equilibrium. 
Because it is impossible to obtain a common absolute height scale for all the pixels in the field of view without additional assumptions \citep{borrero2019}, we used the optical depth of the solar continuum at 500 nm ($\tau_{500}$) as a proxy for geometrical height. The spectral lines observed by Hinode SOT/SP are sensitive to atmospheric layers from $\log\tau_{500}=0$ (defined as the base of the photosphere) to $\log\tau_{500}=-2$ (upper-mid photosphere), corresponding to a height difference of approximately 300\,km in semi-empirical models of the average quiet Sun \citep[e.g.][]{falc}.

The peak sensitivity of these lines to the velocity and the magnetic field is between $\log\tau_{500} = -1$ and $\log\tau_{500} = -1.5$ \citep[e.g.][]{2014A&A...572A..54B}. In the following, we will use $\log\tau$ instead of $\log\tau_{500}$ for brevity.
The SIR code calculates the emergent intensity to be fitted to the observed one in units of the mean quiet Sun continuum, which is calculated from a reference model atmosphere \citep[most often the HSRA,][]{1971SoPh...18..347G}. Since we don't have a proper flux calibration source covering these datasets we normalize the spectra to the mean continuum intensity of the observed quiet Sun for any given dataset; although this implies that absolute changes in temperature cannot be detected, changes in the slope of the temperature stratification can be determined from the change in the shape of spectral line profiles. SIR uses the LTE approximation to describe the ionization and excitation state of the atmospheric plasma and calculate polarized absorption and emission terms. \cite{Smitha_2020} highlighted how non-LTE might influence the formation of neutral iron lines, like the ones observed by Hinode SOT/SP. Due to complex atomic models for neutral iron, performing a rigorous non-LTE inversion of these lines is computationally unfeasible for large datasets. Therefore, we perform the inversions in LTE, possibly losing on absolute accuracy, but hoping to capture the general trends in the temporal changes of the temperature gradient. The non-LTE effects do not influence the inference of the line-of-sight velocities.

Depth-stratified spectropolarimetric inversion codes like SIR allow a flexible choice of the complexity of atmospheric models, allowing the user to specify the number of depth ``nodes'' for each physical quantity, i.e. the number of fixed points equally spaced in $\log \tau$ that are treated as free parameters. The best-fit atmosphere is obtained by perturbing values in the nodes and then interpolating the perturbation in the rest of the atmosphere using a cubic spline. Our inversions used 5 nodes in temperature, 3 in line-of-sight velocity, 2 in magnetic field strength, and one node each for the azimuth and inclination of the magnetic field. For completeness, we also include a single node in micro-turbulent velocity, but due to degeneracies (see below), the inversion code does not vary the micro turbulent velocity significantly from the initial starting value of 0 km/s. The reason for choosing a simple configuration for the magnetic field is that we are focused on determining thermodynamic properties in the non-magnetic atmosphere. To this end, a simple determination of the presence of a magnetic field is sufficient to identify and exclude strongly magnetized pixels.

Different node configurations result in different best-fit atmospheres due to the non-linearity of the inversion process and the degeneracy between the inferred parameters \citep[see e.g. the discusssion in][]{2006A&A...456.1159M}, for example the microturbulent velocity and temperature which both broaden the spectral line. Furthermore, inversion codes often get stuck in the local minima of the minimization metric and tend to be heavily influenced by noise in the observations. As a result, the spatial ($x,y$) distribution of the inferred parameters tends to be discontinuous. To obtain smoother spatial distributions of the inferred parameters, we used the strategy described in \citet{snapi} where the inferred atmospheres are smoothed by a 3x3 pixel median average in the $x,y$ domain, and the result is used as an initial guess for another inversion run. We found that this approach had the best results when repeated twice. In other words, the observed spectra were originally inverted with a single starting atmosphere, a second time with the spatially filtered output of the first inversion as an initial guess, and the third time with the spatially filtered output of the second inversion as an initial guess. This method improved both the quality of the fits and the spatial coherence of parameter maps.

\begin{figure}
    \centering
    \includegraphics[width=0.5\textwidth,scale=1]{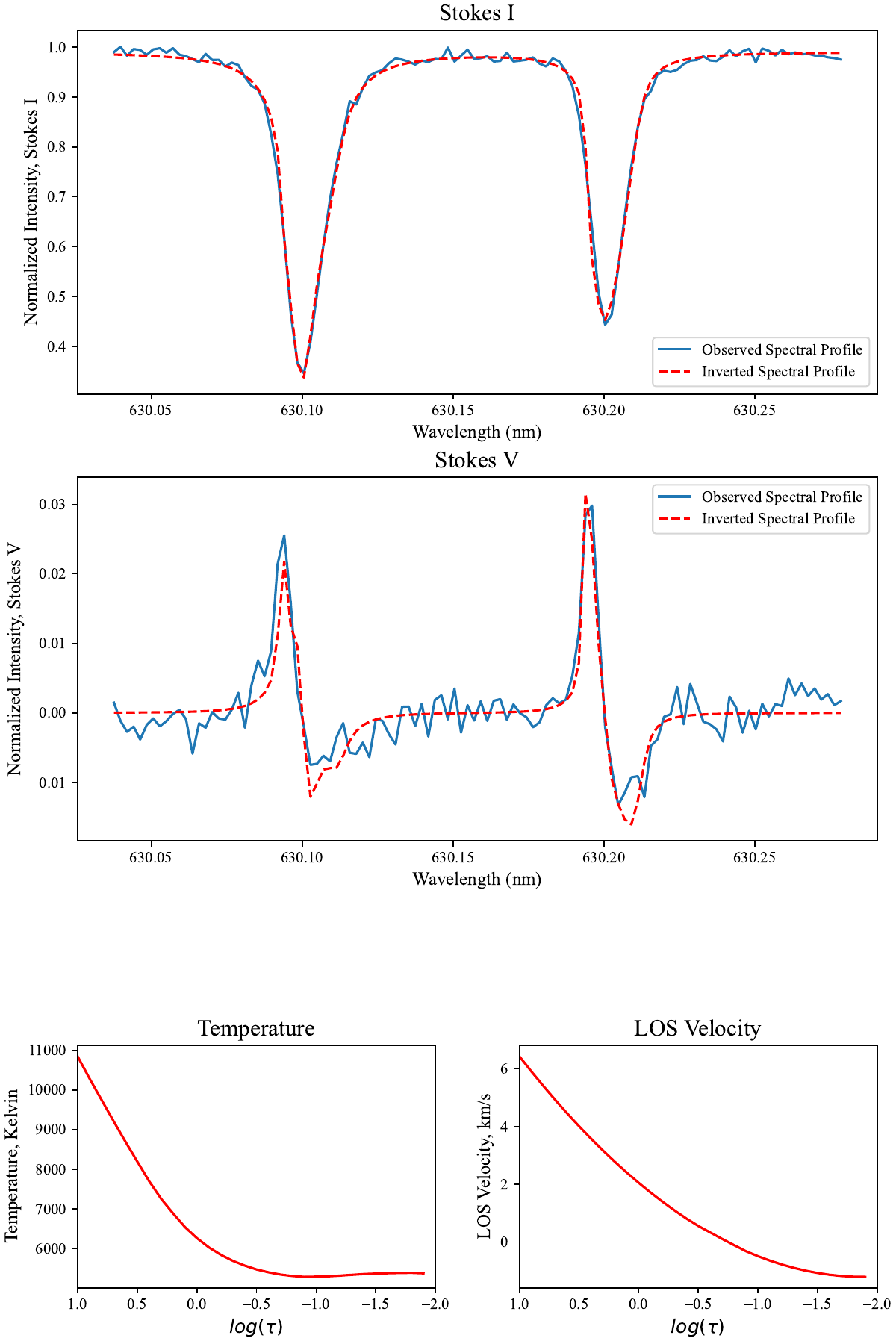}
    \caption{A single example inverted spectra. The top two panels show the observed Stokes I and V components (blue lines), as well as the best fit from a SIR inversion (red dashed lines). The bottom panels show the run of the inferred temperature and line-of-sight velocity with optical depth. The magnetic field strength inferred for this pixel is 80 G at $\log\tau = 1$ (not shown).}
    \label{fig:example_fits}
\end{figure}

We inverted the whole 165 $\times$ 165-pixel ($51.8" \times 24.5"$) region from each data set, amounting to over 1.3 million spectra over the 49 datasets. Figure \ref{fig:example_fits} shows an example spectral profile of the Fe I doublet observed by Hinode/SP, both in intensity and circular polarization (Stokes I and Stokes V, blue lines), as well as the resulting fits from a SIR inversion (red dashed lines). The bottom panels show the run of temperature and line-of-sight velocity with optical depth in the model atmosphere. The line-of-sight magnetic field strength for this pixel amounted to 80 Gauss at $\log\tau = -1.0$.

\subsection{Clustering Intergranules and Granules} \label{subsec:clustering}

% kpr
Due to the physical differences between granular cells and intergranular lanes, we observe large differences in the atmospheric models inferred from different spectra within a single dataset. At the base of the photosphere, granules are brighter, hotter upwellings of plasma while intergranules display cooler, rapid downflows. It is thus likely that if cycle-related, systematic changes in the photosphere do exist, they are unequally expressed in granular and intergranular structures. For example, \citet{2021A&A...652A.103B} found differences in the size and filling factor of granules over the solar cycle. These changes could also affect the stratification with height of the granular and intergranular thermodynamical parameters.

To separate spectra in groups, we used k-means clustering, a technique commonly used in signal processing and image segmentation applications \citep{macqueen1967some}. In general terms, k-means clustering takes a number of clusters (`k') as an input and groups the data by minimizing the intra-cluster distance across all dimensions and in all clusters simultaneously.
%, which is equivalent to optimizing the intra-cluster over all clusters. 
For our case, we applied k-means clustering to the spatially resolved spectral profiles in each dataset. The high resolution of SP/Hinode allows us to unveil clear differences in the profile shapes of spectra originating from granules or intergranular lanes, and the k-means algorithm is able to well separate the intrinsic differences in the spectra. The k-means clustering method has often been used in solar physics because it is a transparent and efficient method to separate high-dimensional data \citep[e.g.][]{2024ApJS..271...24S}. 

\begin{figure}
    \centering
    \includegraphics[width=0.5\textwidth,scale=1]{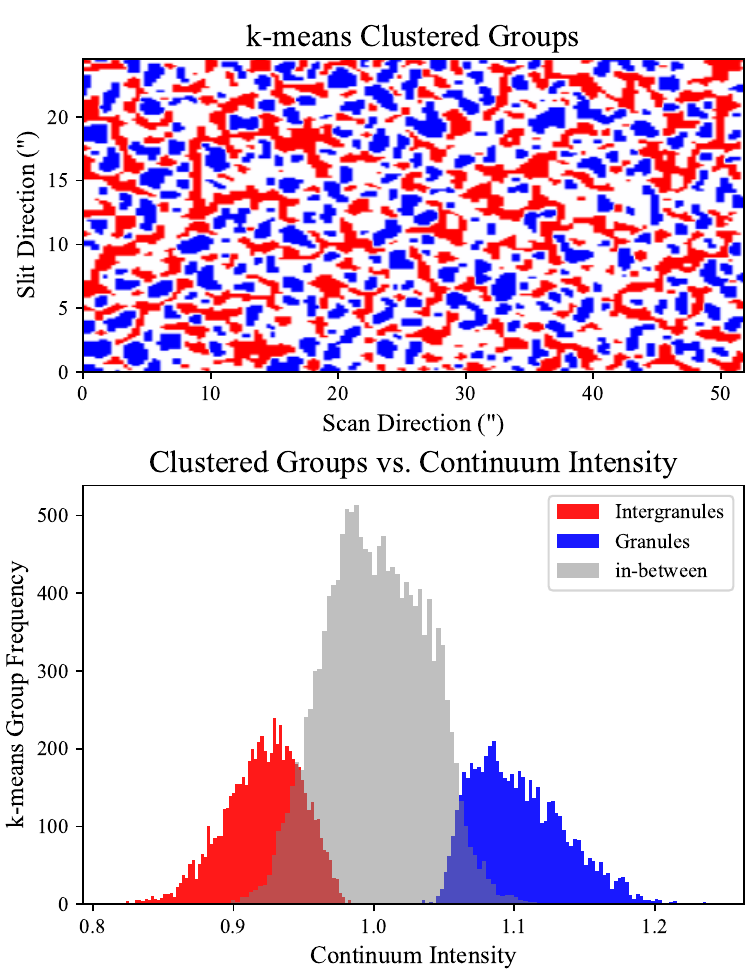}
    \caption{Example clustering of the 2008/12/17 04:59:05 dataset using k-means clustering on the full spectral profile of each pixel. The top panel shows the spatial distribution of granules (blue) and intergranules (red). The bottom panel displays the histograms of continuum intensity in each group, including the intermediate cluster. There is almost no overlap between the continuum intensity of granules vs. intergranules.}
    \label{fig:clustering}
\end{figure}

We found that the observed spectra were best clustered into granules and intergranules when using three groups. With only two groups, the boundaries between granules and intergranular lanes caused the k-means algorithm to often 
create a large overlap between the distributions of fundamental parameters such as continuum intensity or line-of-sight velocity. With three groups, instead, the k-means algorithm well-separates out granules and intergranular lanes with a border of pixels in between. 
The results of the clustering for one of the datasets are seen in Figure \ref{fig:clustering}. 

Pixels clustered as granules represent, on average, 22 \% of all pixels (about 6000 pixels/dataset). As expected, these pixels tend to lie at the center of brighter areas observed in the continuum intensity and have line-of-sight velocities directed toward the observer. Pixels clustered as intergranules represent, on average, 26 \% of all pixels (about 7000 pixels/dataset). The inverted atmospheres of intergranules also match our expectations for the physical nature of intergranules: they display line-of-sight velocities directed away from the observer, higher magnetic fields (see Section \ref{subsec:filtering}), and cooler temperatures deep in the photosphere. We note that more than half of the pixels are classified as ``boundaries'' and are not further considered; our analysis thus concerns only the areas where convective properties are most pronounced. Of relevance for our later analysis, the number of pixels classified as granules or intergranules vary only by a few percent within datasets, without any obvious correlation with either the solar cycle or the changes in any quantities.
% removing this March 28th 2025:
%The number of pixels in each class is noted in Table \ref{table:datasets}}.

\subsection{Filtering of Magnetic Pixels} \label{subsec:filtering}

\begin{figure}
    \centering
    \includegraphics[width=0.5\textwidth,scale=1]{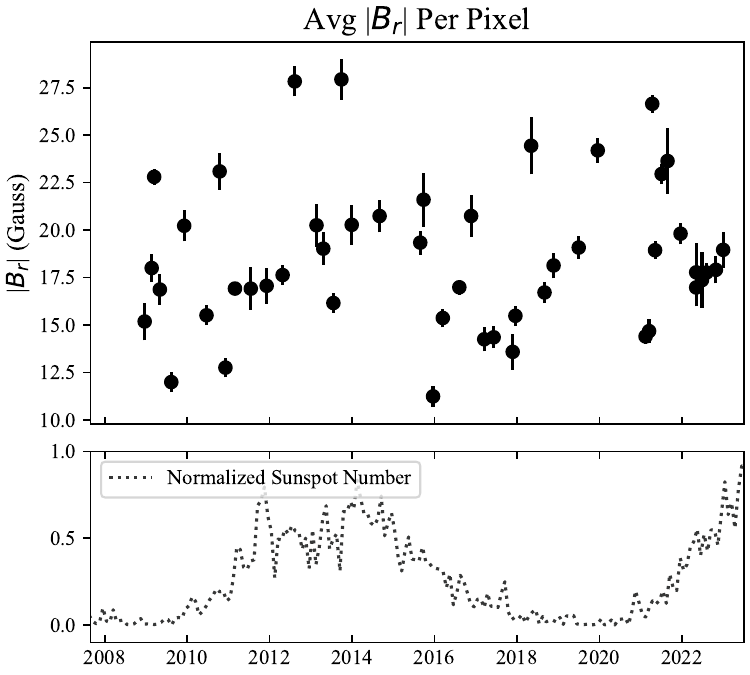}
    \caption{The average magnetic flux per pixel in the remaining pixels after thresholding those with magnetic fields above 200 Gauss. Plotted below the time series is the International Sunspot Number for comparison.}
    \label{fig:B_vs_time}
\end{figure}

Despite specifically choosing datasets targeting the quiet Sun, all of them included at least some pixels with detectable magnetic fields. Because this study focuses on non-magnetic features, spectra with an inverted magnetic field strength above 200 Gauss were masked out and not considered in the final analysis. The result was that up to 5\% of pixels were excluded from each dataset: of these, only a small fraction, about one-tenth, were originally clustered as granules, while the rest were approximately evenly split between the intergranule and boundary clusters. The average flux density in each scan after removing the magnetic pixels was in the 15--25 G range and, importantly, we observed no correlation with the solar cycle (Figure \ref{fig:B_vs_time}). 
The time series of average magnetic fluxes shows no correlation with the solar cycle, as demonstrated by the sunspot number plotted below the figure. The error bars on the plot are calculated by the method described in Section \ref{subsec:temp_diff}.

\section{Results} \label{sec:results}

\begin{figure}
    \centering
    \includegraphics[width=1\textwidth,scale=1.0]{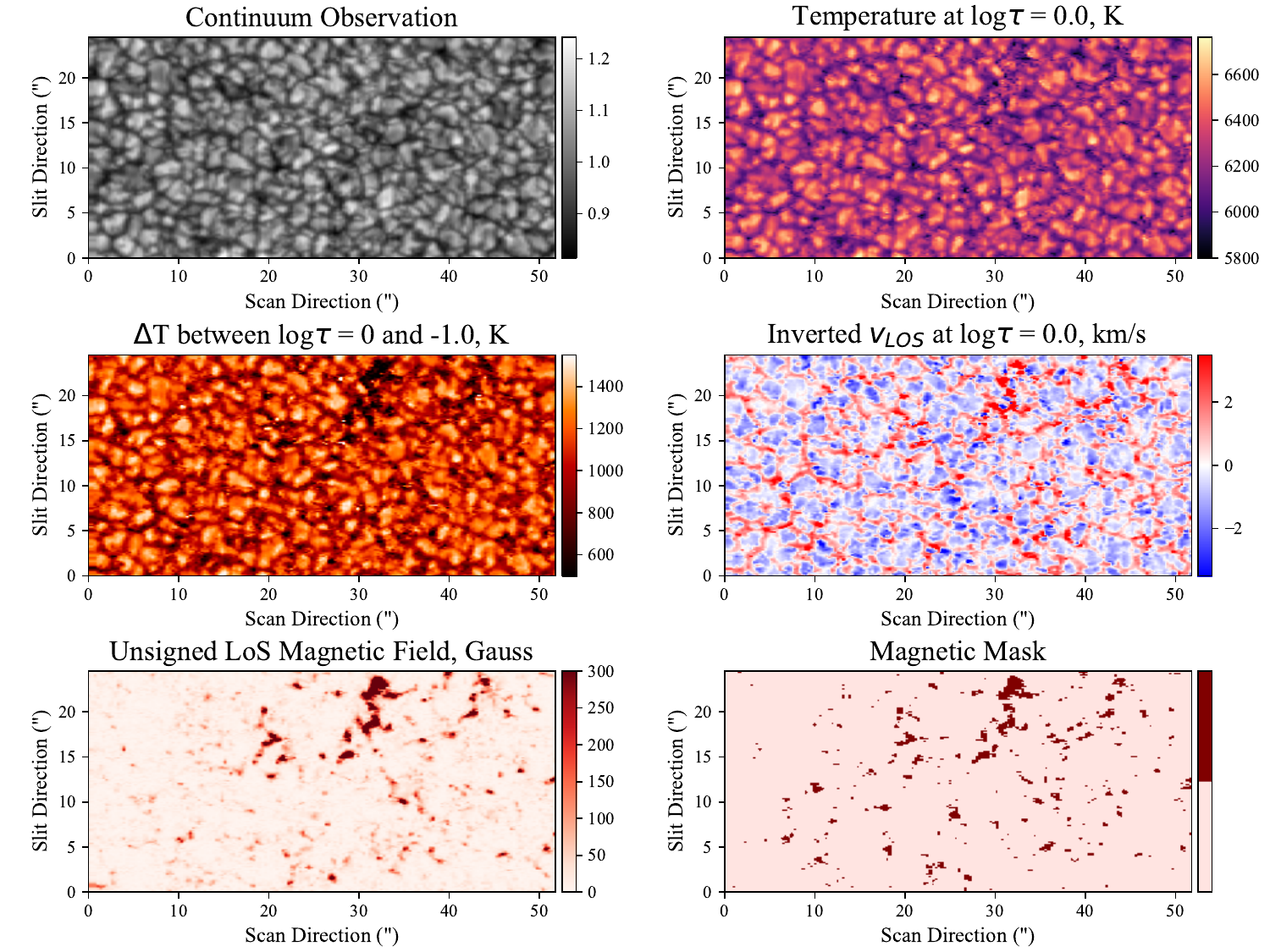}
    \caption{Example of inverted results for the same dataset of Fig. \ref{fig:clustering}. Top left: Continuum at 630\,nm, normalized as described in the text. Top right: the inferred temperature at the $\log\tau=0$ layer. Middle left: the difference between the inverted temperature at two depths: $\log\tau=0$ and $\log\tau=-1$. The very dark areas (shallower temperature gradient) correspond to regions of stronger magnetic fields. Middle right: Line-of-sight velocity field at the surface ($\log\tau =0)$. Bottom left: Line-of-sight magnetic (LOS) field, in Gauss at $\log\tau = -1$. Bottom right: The mask of magnetic pixels excluded from the analysis for this dataset}.
    \label{fig:inversion}
\end{figure}

By inverting and clustering the spectral profiles for all 49 datasets with the process described above, we obtain the distribution of atmospheric profiles in granules and intergranules partially spanning solar cycles 24 and 25. Figure \ref{fig:inversion} shows an example of the inversion results for the same dataset of Fig. \ref{fig:clustering}. The top two panels clearly show the correspondence between continuum intensity and temperature at the surface ($\log\tau = 0$) from the SIR inversion. The average temperature at $\log \tau = 0$ is 6390 K which is set by the corresponding value in the HSRA model atmosphere. In the temperature map, several areas exhibit a ``salt and pepper'' discontinuous appearance; most of these pixels correspond to relatively strong magnetic features, as depicted in the bottom two panels. The middle panels (temperature gradient and line of sight velocities) give an example of the results discussed in the next subsections. 
% removing this, repeat of previous sentence.
%In the following, we focus on the stratification of the temperature and line-of-sight velocity. 

\begin{figure}[h]
    \centering
    \includegraphics[width=150mm,scale=1]{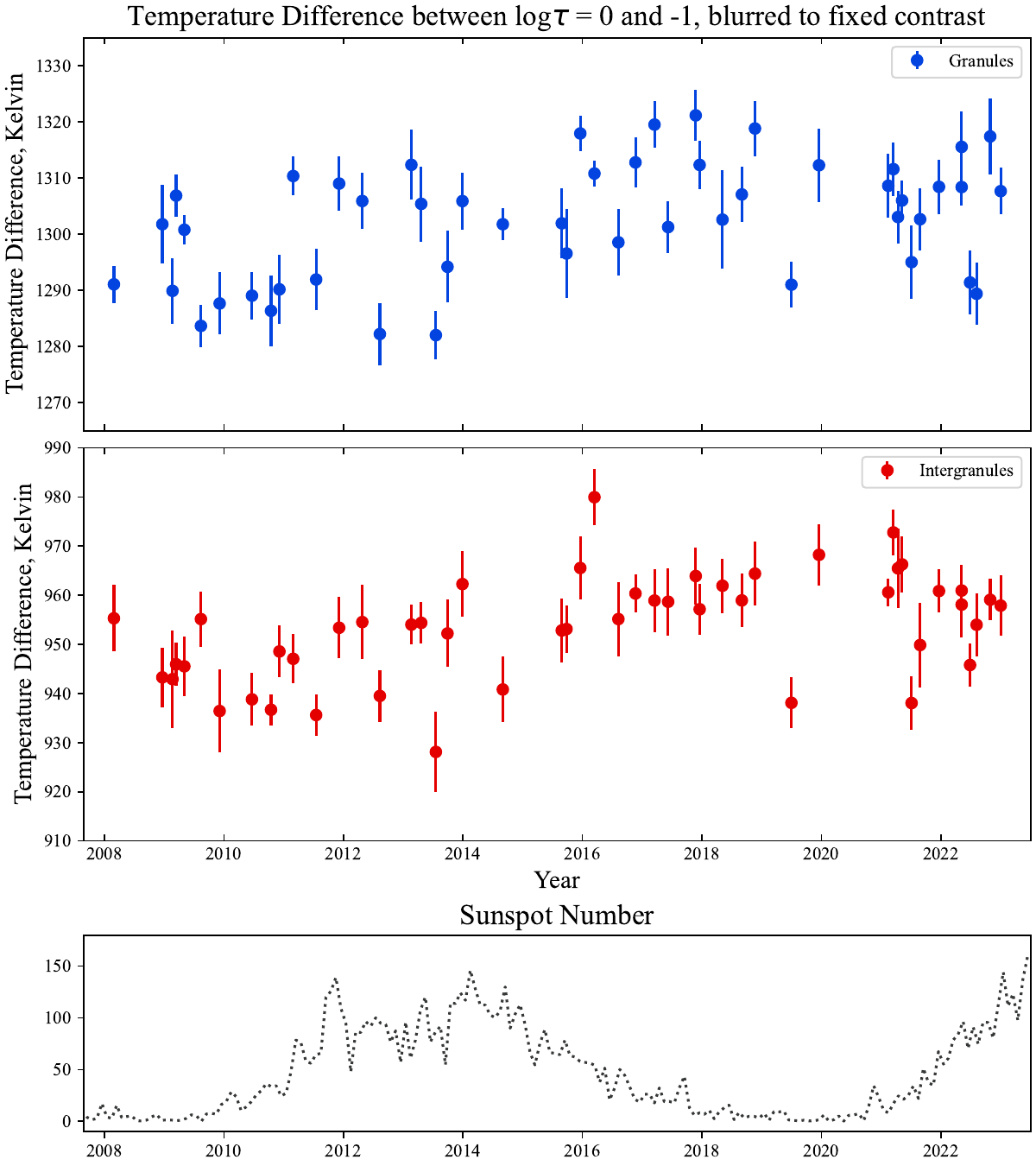}
    \caption{Temporal evolution of the spatially averaged temperature difference between optical depths of $\log\tau = 0$ and -1 in spectra clustered as granules (top) and intergranules (middle). As in Fig. \ref{fig:B_vs_time}, the bottom panel shows the International Sunspot Number for comparison.}
    \label{fig:delta_T_0_-1}
\end{figure}

\subsection{Temperature Difference}
\label{subsec:temp_diff}

%{\bf As discussed in the introduction, } energy transport 
The temperature variation and local gradients are measures of radiative energy flux in the photosphere, and are commonly analyzed quantities in both inversions of observations and models of the solar photosphere \citep{2014ApJ...788..151C, 2016A&A...595A..71F}. We define the temperature difference $\Delta T_{(0,-1)}$ as the difference between temperatures at optical depths $\log\tau =0$ and $\log\tau = -1$, i.e. between the solar surface and the mid-photosphere:

\begin{equation}
\Delta T_{(0,-1)} = T(\log \tau = 0) - T(\log \tau = -1)
\end{equation}
Both Fe I lines used in this work are sensitive to physical conditions in this range of heights \citep{grec_measuring_2010}, corresponding to a difference of 150--200\,km of geometrical height, according to semi-empirical model atmospheres. 

Fig. \ref{fig:delta_T_0_-1} shows the temporal variations of the $\Delta T_{(0,-1)}$, averaged separately over granules and intergranules within each dataset. The error bars on the inferred quantities are calculated from the standard deviation of the mean of 9 subfields of each dataset. In other words, we split each inverted dataset into a 3 $\times$ 3 grid of 55 $\times$ 55 spectra and calculate the spatial mean quantities in each class in each subfield. We then take the standard deviation of those means to obtain an overall uncertainty for each physical parameter. By doing this, we estimate the variation in the quantities due to random noise but also the inherent spatial variations in the number or distribution of granules and intergranules within the Hinode quiet Sun datasets.
The typical error within a single dataset is around 5\,K, which is small with respect to the absolute (systematic) errors caused by the simplified physical model, uncertainties in opacity and atomic parameters, assumption of LTE, and imperfect knowledge of instrumental properties.

\begin{figure}[h]
    \centering
    \includegraphics[width=150mm,scale=1]{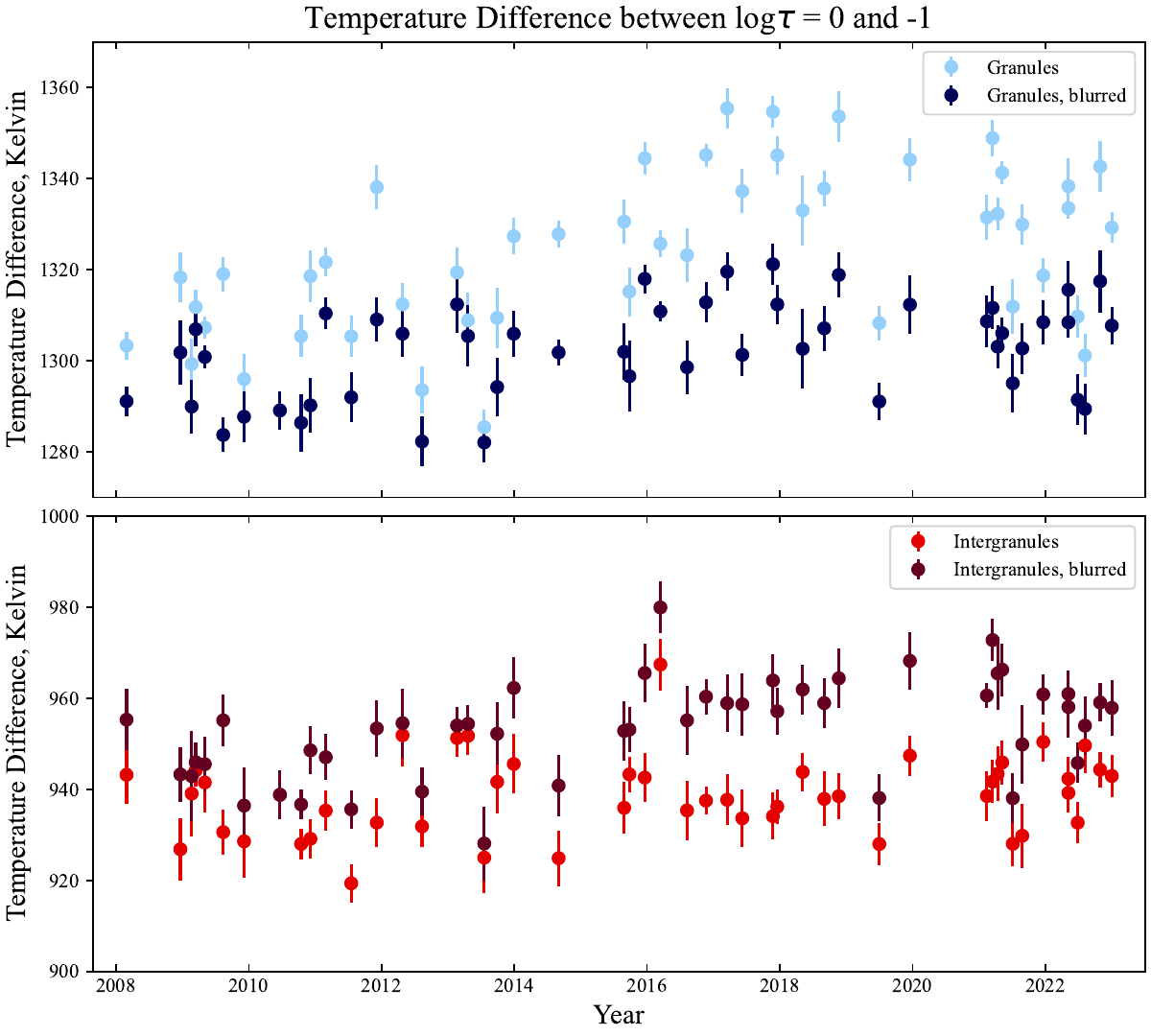}
    \caption{Comparison of original inverted temperature difference (lighter colors) and where the observed data has been blurred to a constant continuum contrast before inverting (darker colors).}
    \label{fig:blurred_vs_unblurred}
\end{figure}

From Fig. \ref{fig:delta_T_0_-1} we see that, on average, the temperature difference in granules is some 400\,K higher than in the intergranules, confirming that granules have a steeper temperature gradient throughout the photosphere. This is clearly manifested in the well-known inverse granulation pattern: in the mid-photosphere, at the convective overturn layers, intergranular lanes are hotter than the granules due to the rapid cooling of the up-flowing parcels of plasma \citep{2006A&A...450..365J, 2007A&A...461.1163C, 2009LRSP....6....2N}. Although these pixels are not represented in the plots of Fig. \ref{fig:delta_T_0_-1}, from the middle left panel of Fig.\,\ref{fig:inversion} we also note that the atmospheric stratification in the magnetic pixels is noticeably different than for the normal convection pattern, with a very shallow temperature gradient (cf. the dark areas in the map).

We find that the temperature difference, $\Delta T_{(0,-1)}$, remains fairly constant over the 15 years, exhibiting a range of temperature differences of about 40\,K over the whole time series. 
%However, due to the large number of datasets analyzed, w
We are able to observe a slight apparent increase in the temperature difference of both granules and intergranules between 2015 and 2021, during the descending phase of Solar Cycle 24. Between 2013 and 2015, there is a slight increase in the temperature difference of both granules and intergranules of approximately 15\,K. For roughly 5 years afterward - from approximately 2015 to 2020 - the temperature difference is roughly constant. Afterward, there appears to be a decrease towards the original value in both granules and intergranules, but more recent datasets are required to determine with certainty. Ideally, an additional declining phase of a solar cycle could be able to confirm these proposed trends, but at present, only one has been observed during the Hinode mission.

We noted above that there are temporal variations in the continuum contrast, presumably due to changes in focus of the Hinode/SP instrument, which were mitigated by blurring all the datasets to a common contrast. To illustrate the effect of even such a small change in the observations on the inversion results, we plot the temperature differences for both the original and blurred datasets in Fig. \ref{fig:blurred_vs_unblurred}. The blurred case tends to cause the temperature difference to decrease in granules and increase in intergranules as there is more spatial mixing of the signals between the granules and intergranules. %regions of high and low temperature differences. 
This also indicates that the absolute temperature differences derived from the inversion process will be highly dependent on the effective spatial resolution of the data set. Having corrected for this largest systematic variation, we believe that the remaining temporal trends are plausibly dominated by solar behavior.

A similar quantity to the difference in temperature between two layers is the local gradient of the temperature at a given optical depth. This quantity has been shown by \citet{2013ApJ...778...27C} to characterize energy transport in the photosphere and the results are shown and analyzed in the Appendix.

\subsection{Line of Sight Velocities}
Height-dependent shifts of the line absorption profile cause asymmetries observed in the emergent Stokes spectrum, which allows the inversion code to recover the line-of-sight (LOS) velocity at different depths in the photosphere. It is reasonable to assume that, as for the temperature, the eventual presence of cycle-related changes in the photospheric structure would leave an imprint on the convective velocities \citep[e.g.][]{1982Natur.297..208L}. Velocity diagnostics retrieved by the inversion code have the advantage of being largely decoupled from the temperature, and do not suffer from any of the usual approximations about the atomic model and ionization state of neutral iron atoms. 
Thus, the retrieved velocities are a somewhat robust measurement of the dynamic conditions in the atmosphere and are largely independent of other changes in the spectral line formation.

\begin{figure}
    \centering
    \includegraphics[width=150mm,scale=1]{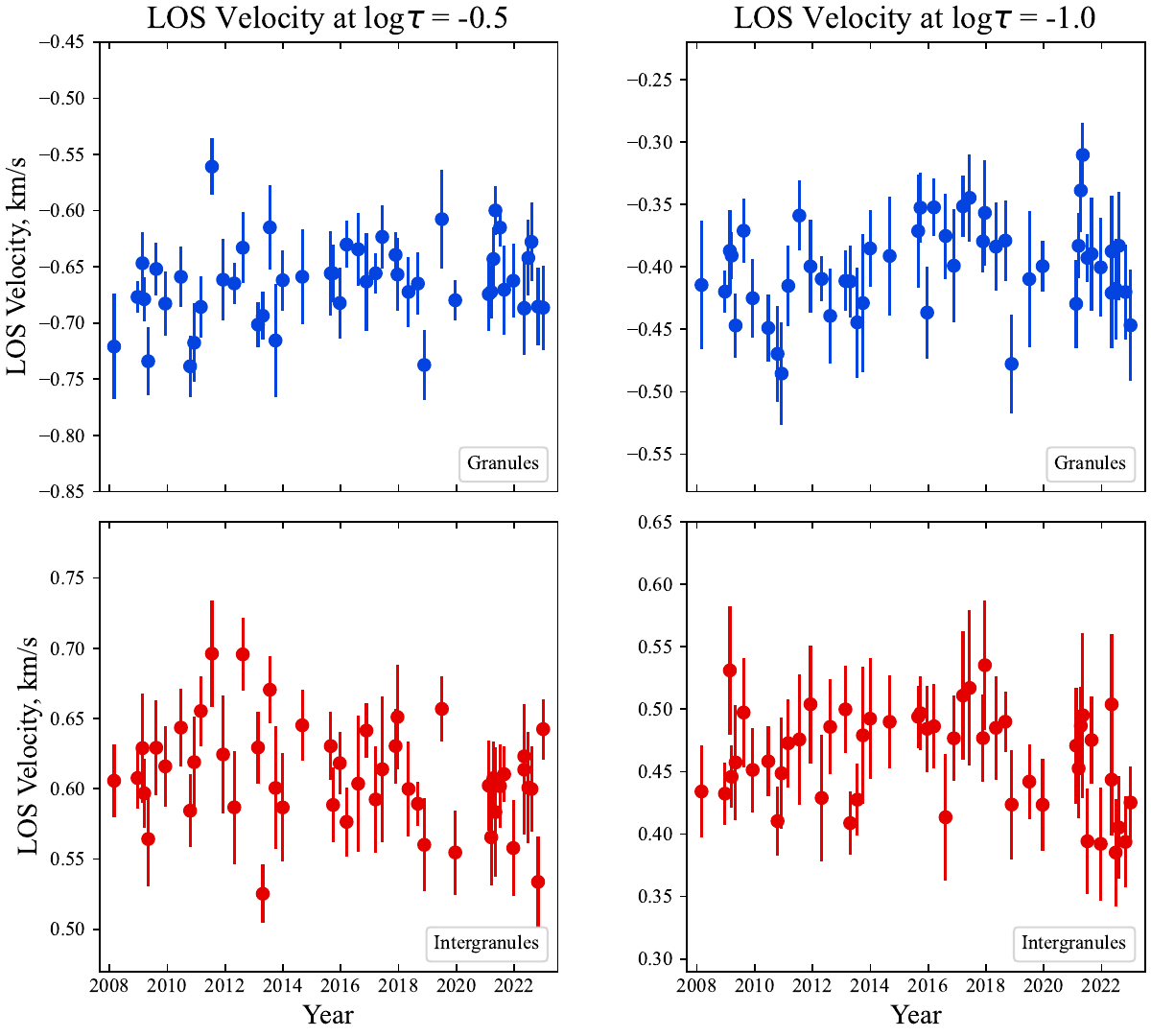}
    \caption{Temporal evolution of the spatially averaged line-of-sight velocity, as obtained from the spectral inversions. Left panels correspond to an optical depth of $\log\tau = -0.5$ (left), while right panels correspond to $\log\tau = -1$ (right). Results for spectra clustered as granules are shown in the top row; intergranules in the bottom row. Negative velocity values correspond to flows directed toward the observer (radial from the Sun as observations are taken at the disk center).}
    \label{fig:los_v}
\end{figure}

Fig.\,\ref{fig:los_v} shows the variation in the retrieved LOS velocity of granules and intergranules at $\log \tau = -0.5$ and $\log \tau = -1$ over the solar cycle. These depths were chosen to be within the range of depths where the lines are most sensitive to thermodynamic properties \citep{grec_measuring_2010, 2014A&A...572A..54B}. Uncertainties were calculated in the same way as described above, and amount to about 0.05 km/s for all features and depths. In order to determine these LOS velocities, we use the wavelength dispersion given by the Hinode FITS headers and set the average velocity over each FOV to 0. Given that each FOV is a sufficiently large area to cover many granules and p-modes, it is reasonable to assume that this change corrects only for instrumental effects and not bulk velocities.

The values depicted in Fig.\,\ref{fig:los_v} are roughly consistent with the reported results in \citet{2017ApJ...836...40O}, and confirm the pattern of deceleration/acceleration of the convective motion inside
granules/intergranules as a function of height, starting from the mid-photosphere.
% KPR - save for later
%Most relevant for our analysis, no clear cycle-related temporal trend is discernible at either optical depth, both for granules and intergranules, 
%at least within our estimated
% KPR - we explain below that this variation is largely due to the p-modes
% although variations up to $\pm 0.05$ km/s are recorded among datasets, maybe due to the relatively small FOV analyzed. 
% KPR - split up the cycle variations from the discussion of their small variations.
%Some previous studies of convective flows in the photosphere also did not uncover any significant variation of LOS velocities in the quiet Sun across several years, e.g.  \citet{2017A&A...598A..99R}. These authors report similar uncertainties of their measurements (of the order of  0.1 km/s), but their year-to-year variations are sensibly lower than what is shown in Fig.\,\ref{fig:los_v}, only of the order of  only $\pm$ 0.02 km/s. 
\citet{2017A&A...598A..99R} measured the temporal variation of the average velocity measured by SDO/HMI, summing over a 300'' $\times$ 300'' FOV (i.e. 70 times greater than our Hinode FOV), and found day-day-day variations of only a few m/sec. This is sensibly lower than the several tens of meters per second that is seen in Fig.\,\ref{fig:los_v}.

It is important to note that our retrieved velocities encompass not only the granular (convective) motions but also include a sampling of the \textit{p}-mode velocities present during the scan. Since the raster scans mix the spatial and temporal information along one axis, it is hard to separate these two components directly from the data \citep[as was done by, e.g.][]{2017A&A...598A..99R, 2017ApJ...836...40O}. The short duration (15 minutes) and limited spatial coverage of our scans imply that there won't be a large number of realizations of individual \textit{p}-mode wave packets such that their signal will fully average out in summed velocities. To assess the possible bias introduced by the rastering process, we used a 1-hour long time sequence of imaging spectroscopy velocities measured with the IBIS instrument in the core of the Fe I 7090 \AA\ line, which forms approximately at an optical depth of $\log\tau = -1.0$  \citep[see][for a complete description of the IBIS data]{2023ApJ...952...58V}. 
%on 25 April 2019  
Using the sequence of 2D maps of the LOS core velocity, we construct synthetic scans of the quiet Sun with the same spatial and temporal sampling as the Hinode/SP data, both for the original measured velocities 
%(akin to what done with the Hinode/SP data), 
%(i.e. like the spectrograph data, mixing the difference velocities) 
and after performing a sub-sonic filter to remove spatio-temporal structures with apparent speeds greater than 7 km/sec, which serve to effectively remove the bulk of the \textit{p}-mode velocities. Extracting multiple synthetic scans from different parts of the overall IBIS data cube and comparing filtered and unfiltered velocity maps, we find that the varying contribution of the \textit{p}-modes introduces an RMS uncertainty of approximately 0.03 km/sec in the average velocity of any given scan. This can account for most of the scatter seen the Fig.\,\ref{fig:los_v}. Importantly, the \textit{p}-modes as sampled have an essentially zero mean and do not bias the value of the averaged velocities.  

% conclusion to velocity section
%\gc{(The following is TBD)} 
Within the limits of these uncertainties, we see no clear trends in the granular or intergranular velocities over the observed interval, nor any correlation with the solar cycle. This is consistent with the findings of \citet{2017A&A...598A..99R} who found no temporal variation in the overall average velocity measured around disk center.

%At first glance, the lack of systematic variations in the LOS convective velocities over the 15 years of Hinode/SP data seems at odds with earlier results, such as those obtained by \cite{1982Natur.297..208L} or \cite{cavallini_long-term_1986}, who both detected cycle-related variations in the shape of different photospheric Fe\,I lines. \gc{However, these authors analyzed only spatially averaged spectral profiles over a quiet Sun area (about $1 \times 2$ arcmin for Livingston \citeyear{1982Natur.297..208L}, and $23 \times 1$ arcsec for Cavallini et al. \citeyear{cavallini_long-term_1986}), without a clean discrimination between magnetic and non-magnetic elements. The reported ``weakening of convection'' could then be ascribed to the presence of a variable number of quiet-Sun magnetic elements entering the average, rather than to a change in the properties of convective cells proper. This could be consistent with the results shown in  Fig. \ref{fig:los_v}.}

\subsection{Bisector Analysis}

%Several previous studies have reported systematic variation in the convective properties over period of several years. 
Older studies, such as the landmark paper of \cite{1982Natur.297..208L}, 
%or \cite{cavallini_long-term_1986}, 
detected cycle-related variations in the shape of photospheric Fe\,I lines over periods of several years, indicating possible variations in convective properties. These studies relied primarily on characterizing the line profiles by the shapes of their bisectors, which capture the profile shifts at a range of selected depths in the wings of the spectral line. 
%This 
Generally speaking, bisectors are a way of examining the height-dependent velocity profiles, and were a common approach before the availability of reliable %algorithms for depth-dependent 
spectropolarimetric inversions. 
%GC ADDED HERE
For example, the position and shape
%curvature 
of photospheric bisectors have been analyzed to study convective velocity properties \citep[e.g.][]{1991A&A...243..244G} or changes in these properties due to varying amounts of magnetic flux \citep[e.g.][]{cavallini_long-term_1986, 1990A&A...231..221B, 1983IAUS..102..149L, 1984ssdp.conf..330L, 1987rfsm.conf...14L}.
%GC TO HERE
 
\begin{figure}
    \centering
    \includegraphics[width=150mm,scale=1]{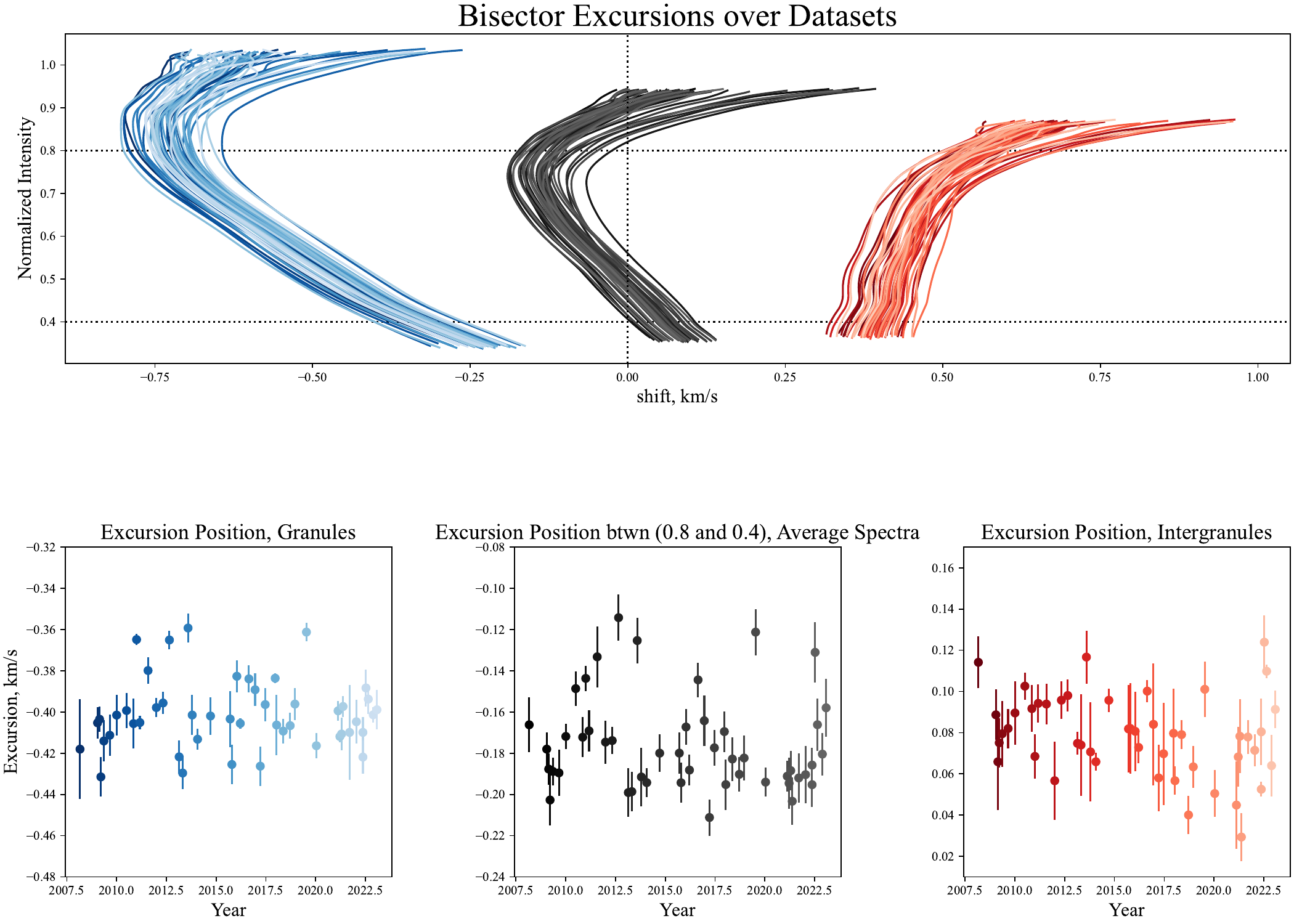}
    \caption{Temporal evolution of bisector velocities over the selected datasets. Dark to light colors indicate early (2008) to late (2022) datasets. Top: calculated bisectors of the average non-magnetic spectra (gray), average granule spectra (blue), and average intergranule spectra (red) from each dataset. Bottom: temporal evolution of the difference between each bisector position at a line depth of 0.8 and 0.4 in each dataset, showing the shape of each bisector. Error bars are calculated the same way as described in section \ref{sec:results}.}
    \label{fig:bisector}
\end{figure}

%Because previous studies have used more simplistic inversion or analysis techniques to search for changes throughout the solar cycle, they do not all have access to inverted quantities like the depth-dependent temperature and line-of-sight velocity analyzed above. 
To directly compare our results to previous studies in a model-independent fashion, 
we calculated the spectral line bisectors of the 630.15\,nm line. 
%We have chosen to calculate the bisector of only the 630.15\,nm line 
We chose this line for the same reasons as \cite{2017ApJ...836...40O}: it is less magnetically sensitive and forms over a wider range of heights than the 630.25\,nm line, and so it provides for a more robust measure of convective and photospheric velocities.
In Fig. \ref{fig:bisector}, top panel, the black curves show the bisector of the average spectral profiles over the years, calculated  
over all the pixels in each dataset, but excluding the magnetic pixels. They clearly show the signature ``C-shape'' of photospheric profiles, which is a result of a combination of the correlated, height-dependent convective and thermal signatures of different photospheric structures  \citep{1981A&A....96..345D}.

% paragraph about findings when seperated out: changes in intergranule bisectors
Using our cluster-defined classes of granules and intergranules, we can also compute the bisectors of the mean spectral lines arising from either of these physical structures. These are shown in blue and red in the top panel of Fig. \ref{fig:bisector} for granules and intergranules, respectively. As with the inversion-derived velocities (Fig. \ref{fig:los_v}), these bisectors indicate that both granular and intergranular flows decelerate with height above the surface (while moving in opposite directions obviously). However, in Fig. \ref{fig:bisector} we can also appreciate that the granules' bisectors display an intrinsic, strong ``C-shape'', which implies a slight upward acceleration in the deeper part of the photosphere, from the surface up to intensities of $\approx$ 0.8 I$_c$, i.e. in the first few tens of km above the surface.

This is consistent with the results of \citet{2017ApJ...836...40O} (who analyzed a single Hinode dataset), although the magnitude of the 
shifts from the nominal rest position, for both granules and intergranules' bisectors, is larger than what those authors found. This might be due to \citet{2017ApJ...836...40O} using a simple velocity cutoff ($\pm$ 0.18 km/sec) to define their granule and intergranule classes, while our clustering-based approach excludes an intermediate boundary class and therefore is weighted to pixels with more extreme velocities.

The top panel of Fig. \ref{fig:bisector} shows that the shape of the average bisectors is very similar throughout the 48 datasets,
but the range of their wavelength positions (vs. the nominal zero) is sensibly larger than what reported in older studies: when looking at different intensity levels we see a spread of $\approx$ 3--4 m\AA~ (0.15 - 0.2 km/s) vs. the reported 1--2 m\AA~ in  \citet{1982Natur.297..208L}. (This spread also appears slightly larger for intergranules than for granules). It is possible that some of this variance is introduced by random variations of the number and properties of granular cells in our relatively small FOV. To investigate any possible secular changes in line shape or bisector shifts, we thus decided to use the bisector ``excursion'' for each dataset, taken as the difference in the wavelength position of the bisector at normalized intensities of 0.8 and 0.4 in the line profile. This allows us to quantify the bisector's shape while removing any absolute shift caused by instrumental or normalization issues. Following \citet{2017ApJ...836...40O}, these two intensity levels correspond to optical depths of $log\tau=-0.25$ and -$1.0$ respectively (corresponding, in turn, to the geometrical height of 50 and 160 km in a reference model atmosphere). Hence, by using this excursion value, 
we should be able to uncover any possible cycle-related variation of the convective flows measured in the middle photosphere.

The magnitude and temporal evolution of the bisectors' excursions are shown in the lower panels of Fig. \ref{fig:bisector}. None of the bisectors, averaged over granules, intergranules, or the full field of view (excluding the magnetic pixels), show any obvious temporal variation correlated with the solar cycle, within an apparently random scatter of about 0.02 km/sec.
This seems at odds with earlier results, such as those obtained by \cite{1982Natur.297..208L} or \cite{cavallini_long-term_1986}, who both detected cycle-related variations in the shape of different photospheric Fe\,I spectral lines.

\section{Discussion}\label{sec:discussion}

From the results presented in Sect. \ref{sec:results}, we tentatively observe faint imprints of the solar cycle in the analyzed thermodynamic properties of the quiet Sun, magnetic field-free photosphere. 
Below, we discuss separately our findings in the context of previous work.
An important caveat is that correction of the varying continuum contrast, described in \ref{subsec:contrast_changes}, has a significant impact on the inferred quantities (recall Fig.\,\ref{fig:blurred_vs_unblurred}). Namely, the temperature difference $\Delta T_{(0,-1)}$ in granules, inferred from the data without the contrast correction, exhibits a strong correlation with the continuum contrast. We put the results on equal footing by introducing additional blurring, but such a process inevitably introduces mixing between the spectra of granules and intergranules. The evidence for that is that the trend, originally visible only in granules, after additional blurring appears both in granules and intergranules. This, in turn, means that the spatial resolution of Hinode/SOT/SP is still insufficient to rigorously resolve the surface granulation for purpose of our study.
Our process of blurring to correct for continuum contrast may imply that the full amplitude of a possible trend is, in fact, larger but is hidden by the limited spatial resolution of the observations. An additional constraint that is placed on our analysis by using lower-than-ideal resolution data is the ability to only reliably cluster data into a single class for granules and intergranules. By using observations with higher resolution, it would be possible to resolve the full dynamics of convective flows, for instance, structures internal to granules that might respond differently to changes throughout the solar cycle.

\subsection{Dynamics}

The average convective velocities at different heights within the low-mid photosphere display the expected amplitudes and gradients, with upflows decelerating from $-0.70$ to $-0.45$ km/s between the $\log \tau =-0.5$ and $\log \tau = -1.0$ layers, and downflows accelerating from $0.5$ to $0.7$ km/s in the same height span (in the opposite direction). These values remain essentially constant throughout the years within an uncertainty of about 0.05 km/s, without any temporal trends or obvious correlation with the sunspot cycle (Fig. \ref{fig:los_v}). This appears consistent with the findings of 
\citet{2017A&A...598A..99R}, although their day-to-day variations were much smaller (few m/s) on the account of their much larger FoV and lower spatial resolution. 

We reach the same conclusion when using the line bisectors as a diagnostics for line-of-sight velocities at different heights in the atmosphere. In particular, to avoid any random variation of absolute shifts given our small FoV, we used the bisector excursions between the normalized intensity values of 0.8 and 0.4 in the averaged line profile as a proxy for the velocity gradients within the mid-photosphere. Again, we do not observe any obvious trend or correlation with the solar cycle, within our uncertainties (Fig. \ref{fig:bisector}).  

The apparent contradiction with ``classical'' results such as those of \cite{1982Natur.297..208L} or \cite{cavallini_long-term_1986}, who detected cycle-related variations in the shape of different photospheric Fe\,I spectral lines, is due to at least two different factors. First, these authors performed spatial averages over quiet Sun areas significantly different than our case, and compared only two different (average) spectral profiles, acquired in two different years. Given the natural scatter from dataset to dataset in our calculated bisectors, we believe that using only a few observations over the cycle is not sufficient to reliably relate the changes in bisectors to the change in solar activity and can be attributed to measurement uncertainty or the inherent scatter. For example, we observe that the differences in the bisectors for the two datasets examined in \citet{cavallini_long-term_1986} are approximately of the same magnitude as the scatter we see among our collection of 49 bisectors. Second, and most important, most of the earlier studies
considered the average spectral profiles without any discrimination between magnetic and non-magnetic pixels in their data (as an extreme case, Livingston \citeyear{1982Natur.297..208L} used disk-integrated data). As it is well known that individual magnetic elements display low-amplitude flows as well as shallow velocity gradients with height \citep[e.g.][among many]{1990A&A...231..221B, 2013ApJ...778...27C}, from our results, it seems plausible that much of the reported spectral changes depend on the presence and amount of magnetic elements within the average profiles rather than cycle-related changes in the basal convective properties.

%  We argue that precisely the variations in the number of these small-scale magnetic elements might be responsible for the bisector variations found by \cite{1982Natur.297..208L} and \cite{cavallini_long-term_1986}. Of course, another reason can be the fundamentally different nature of the instruments used: spectral resolutions in the earlier studies are of the order of $10^6$, while Hinode/SOT/SP spectral resolution is around $2\times10^5$, which is further decreased due to undersampling.

\subsection{Thermal structure}

%Regarding the thermal structure, 
Our analysis of the temperature structure in the non-magnetic quiet atmosphere provided more ambiguous results. We confirm that granules experience a stronger variation of temperature with height than the intergranules: the temperature difference between the solar surface ($\log\tau =0$) and the mid-photosphere ($\log\tau =-1.0$) 
is about 400 K larger for the former (Fig. \ref{fig:delta_T_0_-1}). 
%The same behavior can be appreciated in the plots of the temperature gradients (per unit of $\log \tau$) in the low photosphere (Fig. \ref{fig:grad_T}). This is consistent with our understanding of convection, and the same behavior is well reproduced in models of the solar photosphere \citep[e.g.][]{2009LRSP....6....2N}. 

%Like for the convective flows, the temporal analysis does not identify any secular variation of the temperature differences, or temperature gradients, for the intergranules. Surprisingly, a 
%slight secular trend is instead observed for the temperature gradients of the granular pixels: they seem to steadily increase from about 2010--2012 to 2017--2018 (so  3--4 years after the cycle maximum) and decline after that, \gc{although} no periodicity is apparent in the plots. The peak-to-peak variation of the temperature gradient (per unit of $\log\tau$) is $\approx$ 75 K at $\log\tau=0$ and $\approx$ 40 K at $\log\tau=-0.5$ (Fig. \ref{fig:grad_T}). In both cases, this corresponds to about 3\% of the absolute gradient at those heights. 

The temporal analysis of temperature gradients revealed a slight secular trend in both granular and intergranular pixels: there appears to be a steady increase from about 2010--2013 to 2015--2017 (3--4 years after the cycle maximum and a decline after that, although no clear periodicity is apparent in the plots. Without additional data covering another declining phase of a solar cycle, attributing these secular trends to cycle-related changes is speculative at best. The peak-to-peak variation of the temperature difference between $\log\tau=0$ and $\log\tau=-0.1$ is $\approx$ 25\,K in granular pixels and $\approx$ 20\,K in intergranular pixels (Fig. \ref{fig:delta_T_0_-1}). In both cases, this corresponds to about 2\% of the absolute difference.

Using the same type of Hinode/SP data, but a different, geometry-based method taking advantage of scans at different distances from the solar limb, \citet{2016A&A...595A..71F} also found a variation with time of the temperature gradient of the quiet Sun in the low-photosphere (up to $\approx$ 60 km). In particular, they uncover a steeper temperature gradient in their 2013 sample (close to solar maximum), vs. that measured in the sample of 2007 (close to solar minimum). This is the same trend we observe in the plots of Figs. \ref{fig:delta_T_0_-1} and \ref{fig:grad_T}, but a direct comparison with our results is difficult for several reasons: not only did \citet{2016A&A...595A..71F} consider only two epochs within the extended temporal interval analyzed in the present paper, but they also did not distinguish between granules and intergranules like we did. Further, their method allows for the retrieval of temperatures at actual geometrical heights, while our results are expressed in terms of optical depth, as is the case for all spectral inversions.\footnote{They also report this result only for the northern hemisphere, and ascribe this to a strong hemispheric asymmetry of the magnetic activity at the solar maximum of cycle 24} 
%given the spread of values that we observe in our much larger sample, we think however that this explanation might be overly simplistic.}
%
%Another caveat worth discussing is that the inversions naturally yield atmospheric structure on the optical depth scale, instead of the geometrical height. 

Due to the intricacies involved in the radiative transfer, converting one scale into another or performing the inversions in geometrical height is possible, but not without additional difficulties \citep[e.g.][]{BorMil2024_QS}. 
An open question would then be whether any of our observed changes (or lack thereof)
%in the temperature differences, 
might be the result of variations in the optical depth scale vs. the geometrical height. To this end, we note that  \citet{2013ApJ...778...27C}, who investigated the atmospheric properties of the quiet Sun using MHD simulations, states that the results are qualitatively in agreement whether using geometrical or optical depth height scales. Given that our strongest observed variations amount to only a few \% (Fig. \ref{fig:grad_T}), is seems reasonable to assume that the optical depth scale remains fairly stable in our case.
% We note that the general stability of the measured temperature 
%differences ($\pm$2--3\%) might indicate that the optical depth scale is reasonably stable. The other possibility is that the optical depth and temperature profiles might change in unison to produce a generally constant temperature difference.
%The mechanisms for producing any such systematic changes are not clear, but a detailed modeling effort that relies on MHD simulations of the quiet Sun photosphere is needed to fully answer this question, following efforts of \citet{2013ApJ...778...27C}. 

The study of \citet{2013ApJ...778...27C} could in principle provide a relevant comparison with our results. The author used magnetohydrodynamic (MHD) simulations of the quiet Sun to show how different amounts of magnetic flux within the simulations' box (so-called environmental flux) can affect the thermodynamical properties of both, magnetic and non-magnetic (convective) structures. 
For example, the presence of stronger and/or denser concentration of magnetic elements progressively inhibits convection, leading to reduced convective flows even in field-free features; this is consistent with the observed reduction of vertical flows in granulation in areas of increasing average magnetic flux \citep[e.g.][]{2012A&A...542A..96K}. Variation in the temperature stratification can be more complex to interpret, as the presence of increasing environmental flux gives rise to two competing effects: the inhibition of convection (which dominates closer to and below the surface) steepens the temperature gradient, while the reduction of opacity (which is most effective at lower optical depths) reduces the temperature gradient. 

The disk-center Hinode/SP datasets used in our study are all very quiet and contain an average magnetic flux of 15--25\,Gauss, comparable to the weakest of the MHD snapshots analyzed by \citet{2013ApJ...778...27C}. For the gas pressures around $log \tau=-1$, which is $4\times10^4\,\rm{dyn/cm^2}$, we estimate the change in the temperature differences due to the increases in the magnetic pressure and resulting decrease of the gas pressure:
\begin{equation}
nk \Delta T = p_{\rm gas} \frac{\Delta T}{T} = \frac{B^2}{8\pi},
\end{equation}
leading to $\Delta T \leq 5\,$K for $B=25$\,Gauss, which is below the temperature changes we found through inversions. Furthermore, this amount of flux remains essentially constant throughout the years 2007--2022; any variation of the ``environmental flux'' within our scans would then be due to an ambient sub-surface magnetic field which varies due to the overall solar cycle. The overall stability of most thermodynamic parameters as retrieved from the inversions would point towards a negligible effect of such, cycle-related, ambient magnetic fields. We note that for the convective flows, the strongest effects would be observed for the {\it absolute} amplitude of the flows closer to the solar surface, a quantity that cannot be extracted reliably from the inversions; however, as clearly shown in \citet{2012A&A...542A..96K}, measures of the flows in slightly upper layers are also well adequate for this analysis, so we are confident in our results. 
The temperature stratification retrieved from the inversions can be better appreciated even at lower heights above the solar surface, and here, our results are somewhat puzzling. The temperature stratification remains essentially constant throughout the years, with a tentative secular increase of the order of 3\%  from 2010 to around 2016. Given that the total number of pixels identified as granules and intergranules remains essentially constant throughout the years, such a result poses questions about the balance of convective flux in the deeper parts of the photosphere, at different times within the solar cycle. Further investigations about possible shortcomings of the inversion or biases introduced by our clustering technique will be necessary to shed light on this topic. 

\section{Summary and Conclusions}
\label{sec:summary}

We have examined an extensive time series of homogeneous Hinode SOT/SP datasets, spanning solar Cycle 24 and the start of Cycle 25, to investigate possible changes, related to the activity cycle, of the basal thermodynamic structure and convective properties of the quiet Sun photosphere. To this end, we used the spectropolarimetric inversion code SIR \citep{ruiz_cobo_inversion_1992}, which returns a depth-stratified atmospheric model to best reproduce any given spectral profile. To our knowledge, this is the first time that long-term variations of convective properties in the quiet Sun are studied using spectral inversions. 

In order to minimize possible biases due to small-scale magnetic activity, we focused on regions of the quiet Sun at disk center, containing an average flux of 15-25 G, a value that did not change throughout the cycle. Additionally, we removed from the sample any pixel with clear magnetic signatures (threshold of 200 G / pixel).  Our analysis is therefore less susceptible to the effects of variable amounts of unresolved magnetic flux that might be included in spatially averaged spectra, even in the quiet Sun. 
%In contrast to spatially unresolved synoptic time series used in previous studies \citep[e.g.][]{lites_hinode_2007, fang_cyclic_2001}, 

%\textbf{We analyzed all datasets at a constant resolution of around 0.3'', the native SOT/SP spatial resolution of the lowest contrast dataset analyzed}.
We analyzed all datasets at their original sampling (0.32" along the slit, 0.15" slit step) and applied a variable Gaussian blurring to match the granular contrast of all datasets.
These high-resolution observations allowed us to clearly identify and separate the two main distinct physical groups of pixels in the quiet Sun: granules and intergranular lanes, and separately study their properties. This approach is similar to the one adopted by \citet{2017ApJ...836...40O}, but our {\it k}-means clustering algorithm provides a much clearer distinction between the two groups and should provide more representative results related to the convective pattern. By clustering the spectra into physically similar groups, we also avoid the non-linear mixing that arises when summing disparate spectra. By analyzing groups of physically similar spectra together, we avoid effects where the atmosphere inferred from an average spectrum is biased when compared to the average of a set of atmospheres inferred from distinct spectra, as shown by \cite{uitenbroek_why_2011}.

In total, we analyzed a total of 49 datasets over the 15 years from 2008 to 2022, each covering a 51.8'' $\times$ 24.5'' FoV, for a grand-total of $\sim 1.3 \times 10^6$ individual spectral profiles. We concentrated on the average thermodynamical properties for all the pixels within the granules and intergranules' groups (around 6000-7000 per group, per dataset) and how they evolve in time. 
Like for any spectro-polarimetric inversion, results obtained with SIR suffer from systematic errors due to the choice of various parameters (see discussion in Sect. \ref{subsec:inversions}), but since it is difficult to obtain a good estimate of the cumulative absolute magnitude of these errors, in our analysis we have focused mostly on the temporal behavior of quantities' {\it differences}. The significant temporal stability of some of the retrieved quantities (e.g. the  temperature differences, shown in Fig.\,\ref{fig:delta_T_0_-1}), indicate that these uncertainties do not have any large, systematic variations or trends. 

Our main result is that we observe very little variation in most thermodynamical parameters of the low, quiet photosphere. Since the ``environmental flux'' in our datasets remain essentially constant throughout the years at the level of $\approx$ 20 G, any varying, large scale sub-surface magnetic field due to the solar cycle does not seem to have any obvious effect on the properties of the surface convection, at disk center. 

We do, however, find a small, secular change of the temperature gradient in the low photosphere
%, but only for the granular structures, 
with an increase of order of 2\% from 2010 to 2018, and an apparent trend downwards after that date. 
%Since the corresponding quantity for the intergranules remain very constant throughout the years, we think that this probably cannot be attributed to instrumental degradation, and must be due to a yet undefined effect. 
The peak of the temperature gradient is observed in 2018, i.e. few years out of phase with the solar cycle; if related to it, the increase may be tied to the poloidal component of the global magnetic field, similar to what is reported in \cite{korpi-lagg_solar-cycle_2022}. Still, given that our data spans only 15 years in total, additional monitoring in the upcoming years will be necessary for a confirmation. 
If related to the solar cycle, this increase might be tied to the poloidal component of the global magnetic field, similar to what is reported in \cite{korpi-lagg_solar-cycle_2022}. 
Further work on this project should also include repeating this analysis on additional quiet Sun datasets at different heliocentric coordinates, while devising a method to comprehensively compare quiet Sun inversions at different disk positions \citep[following e.g.][]{Trelles2023QS}. 

More generally speaking, additional modern MHD simulations should be investigated to understand any physical explanations and implications of the possible changes we report. For instance, \cite{rempel_small-scale_2018} found that the strength and depth of recirculation in simulations affect the observed structure of the magnetic field and granulation flows, while \citet{Bhatia_SSD_I} showed that the inclusion of small-scale-dynamo in the simulations of surface convection yields detectable changes in the structure of the photosphere for a large range of stellar parameters.

\begin{acknowledgments} 
\label{sec:acknowledgments}
We thank Rebecca Centeno Elliott for the assistance in selecting and preparing Hinode data, for help during the inversion process, and for providing useful feedback throughout the project. We also thank Tom Bogdan for a critical reading of the manuscript. We are grateful for the insightful comments and suggestions from an anonymous referee, which significantly improved the manuscript and the presented results.

Data for this project is provided by the Hinode mission and is publicly available from the Community Spectropolarimetric Analysis Center (\href{https://csac.hao.ucar.edu/sp_data.php}{10.5065/D6T151QF}). Hinode is a Japanese mission developed and launched by ISAS/JAXA, collaborating with NAOJ as a domestic partner, NASA and STFC (UK) as international partners. Scientific operation of the Hinode mission is conducted by the Hinode science team organized at ISAS/JAXA. This team mainly consists of scientists from institutes in the partner countries. Support for the post-launch operation is provided by JAXA and NAOJ(Japan), STFC (U.K.), NASA, ESA, and NSC (Norway).

J.W.C. acknowledges the George Ellery Hale Fellowship of the University of Colorado Boulder for financial support during this project. IM acknowledges the funding provided by the Ministry of Science, Technological Development and Innovation of the Republic of Serbia through the contract 451-03-66/2024-03/200104.

The National Solar Observatory is operated by the Association of Universities for Research in Astronomy, Inc. (AURA), under cooperative agreement AST-1400450 with the US National Science Foundation. This work utilized the RMACC Summit supercomputer, a joint effort of the University of Colorado Boulder and Colorado State University, supported by the National Science Foundation (awards ACI-1532235 and ACI-1532236), the University of Colorado Boulder, and Colorado State University. 

We acknowledge the community effort devoted to the development of the following open-source packages that were used in this work: NumPy (\url{numpy.org}), \citep[][]{harris2020array}, Matplotlib (\url{matplotlib.org}), \citep[][]{Hunter:2007}, SciPy (\url{scipy.org}), \citep[][]{2020SciPy-NMeth}, and Astropy (\url{astropy.org}), \citep{astropy:2013, astropy:2018, astropy:2022}.

This research has made use of NASA's Astrophysics Data System Bibliographic Services, funded by NASA under Cooperative Agreement 80NSSC21M00561.
\end{acknowledgments}

\facility{Hinode}

%\gc{
%Some things we should not forget / discuss:
%\begin{itemize}
%    \item  Ballot+ 2021 have a lot of discussion about the changing physical size of pixels throughout the year due to Earth's orbit. We should say somewhere that given how we identify granules vs intergranules this probably will not matter much. Can we state this? Maybe give the number of granular vs IG pixels in time ? But, would the temperature gradient be affected? THIS NEEDS TO BE DISCUSSED EITHER IN SECT 2.1 OR 2.4
%\item In the Intro we (I) refrained from discussing changes of seismic parameters. Should we go there? I don't understand much of their results, especially because they are global by definition - how do they avoid magnetic elements? I WOULD SKIP THIS AND PRAY THE REFEREE DOES NOT COMPLAIN
%\item We also do not talk about the work done with Hinode about varying of magnetic elements in the internetwork (inside super-granular cells). Do we need it? Can we put in the conclusion? Same for the "stellar overtones". The Intro was already plenty long; maybe we can say something in the Conclusions?
%\item We'll have to debate the issue of finding a gradient (or not) in optical depth scale rather than geometrical height. Not sure where this will lead

%\end{itemize}
%}

\appendix
\label{sec:appendix}

\section{Local Temperature Gradient}

In addition to the temperature difference between two heights in the atmosphere, we also analyze the local slope (gradient) of the temperature stratification which, as shown in \citet{2013ApJ...778...27C}, is closely related to the energy flux throughout the photosphere. Further, the local gradient of temperature makes better use of our depth-stratified inversion: because SIR works by fitting a spline through multiple nodes, the slopes of the temperature at different depths are well-behaved and descriptive of how the temperature profiles are changing. 

\begin{figure} [ht]
    \centering
    \includegraphics[width=150mm,scale=1]{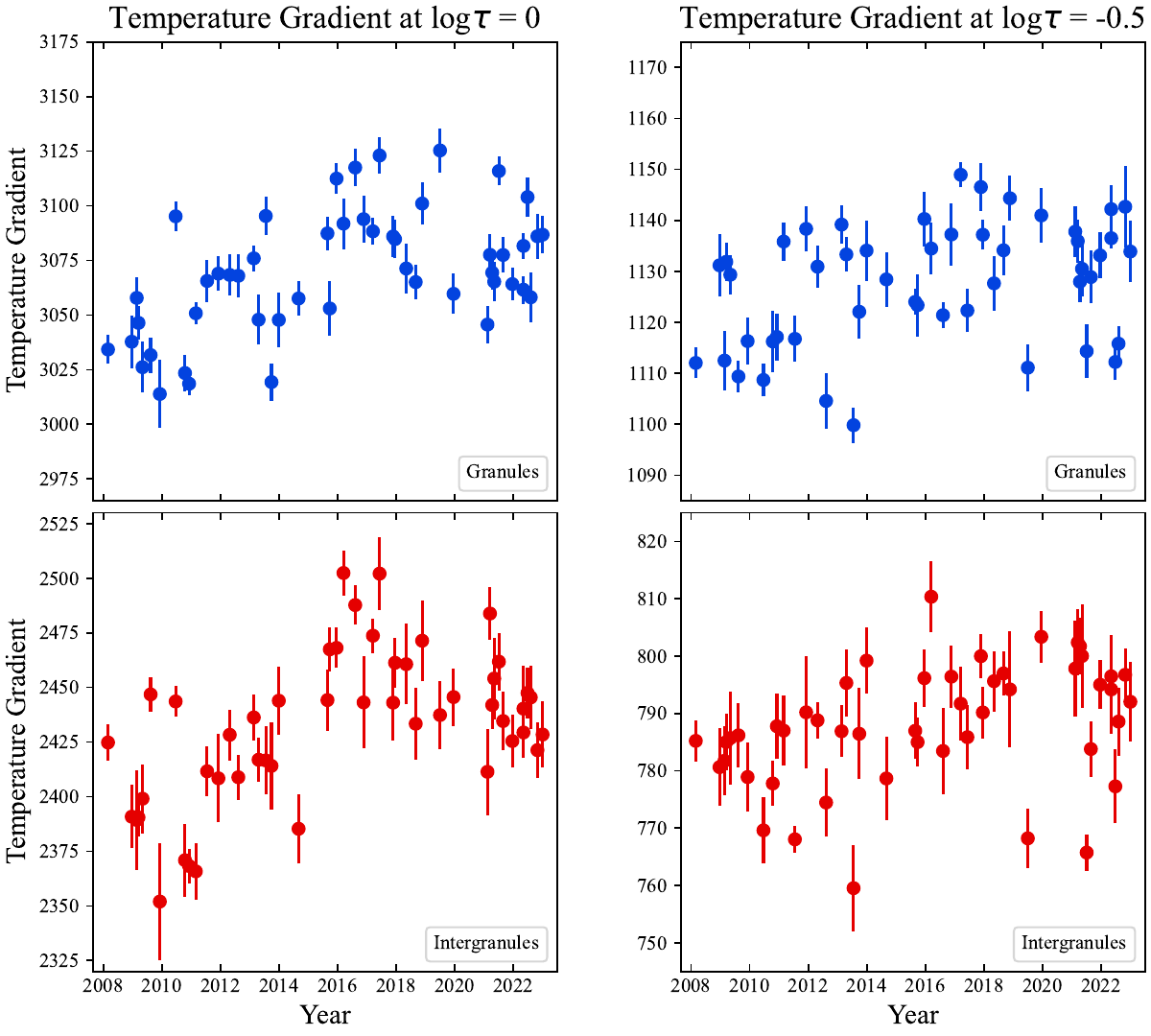}
    \caption{Spatially averaged local temperature gradient in spectra clustered as granules (top) and intergranules (bottom) within each dataset. Left panels refer to an optical depth of $\log\tau = 0$; right panels to $\log\tau = -0.5$. Units on the y-axes are K/unit of $\log\tau$; they are kept equal for plots at the same optical depth (column).}
    \label{fig:grad_T}
\end{figure}

Figure\,\ref{fig:grad_T} shows the local temperature gradients at $\log\tau = 0$ and $\log\tau = -0.5$, in K per unit of $\log\tau$. The gradients are calculated by fitting the temperature curve over a $[-0.1,+0.1]$ interval in $\log\tau$ around the central point, and the error bars are determined with the same method described in the Sec. \ref{subsec:temp_diff}. 
We note that the two chosen central values of optical depth represent atmospheric heights of roughly 0 and 60-80 km \citep[cf.][]{2017ApJ...836...40O}, similar to the heights analyzed in \citet{2013ApJ...778...27C} and  \citet{2016A&A...595A..71F}.
We also note that the larger error bars seen in the left panels with respect to the right ones are a result of the loss of sensitivity of the spectral lines to layers deeper than $\log\tau=0$; hence, although the absolute temperature is well constrained, the local slope is less well determined. 

% https://iopscience.iop.org/article/10.1088/0004-637X/778/1/27/pdf

From the left two panels of Fig.\,\ref{fig:grad_T} we see that the local temperature gradient at $\log\tau=0$ is much higher in granules than intergranules, by about 650 K/log $\tau$. The temperature gradient in intergranules displays a larger variance, as indicated by the larger error bars, 
but both curves show a scatter of $\pm 50$ K/log $\tau$ over the 15 years. 
In both granules and intergranules, we again note a temporal pattern similar to that of Fig.\,\ref{fig:delta_T_0_-1}, with a tentative peak in both quantities in the 2016--2018 time frame.
%{\bf tentative} peak in the 2016--2018 time frame. No clear temporal pattern is discernible for the local temperature gradient at $\log\tau=0$ in the intergranules, with a fairly constant value, especially from 2011 onwards. 

The right two panels of Fig.\,\ref{fig:grad_T}, representing values at $\log\tau=-0.5$, show similar qualitative behavior to the left panels and to what depicted in Fig.\,\ref{fig:delta_T_0_-1}: the temperature gradient at $\log\tau = -0.5$ for granules is about 350 K/$\log\tau$ larger than for intergranules. 
The tentative temporal trends seen in the left panels of Fig.\,\ref{fig:grad_T} and in Fig.\,\ref{fig:delta_T_0_-1} may be present in the temperature gradient at $\log\tau=-0.5$, but the amplitude of temporal trends is smaller and could be more hidden by the natural variation of temperature gradients between datasets.
%the latter shows a remarkably stable pattern throughout the 15 years analyzed, with excursions limited to $\sim 10 $\,K/$\log\tau$; and the temperature gradient in granules shows an increasing trend from 2010 until 2018 (the declining phase of the solar activity), and an apparent decrease afterward. 

\bibliography{bibliography}

\begin{thebibliography}{}
\expandafter\ifx\csname natexlab\endcsname\relax\def\natexlab#1{#1}\fi
\providecommand{\url}[1]{\href{#1}{#1}}
\providecommand{\dodoi}[1]{doi:~\href{http://doi.org/#1}{\nolinkurl{#1}}}
\providecommand{\doeprint}[1]{\href{http://ascl.net/#1}{\nolinkurl{http://ascl.net/#1}}}
\providecommand{\doarXiv}[1]{\href{https://arxiv.org/abs/#1}{\nolinkurl{https://arxiv.org/abs/#1}}}

\bibitem[{{Astropy Collaboration} {et~al.}(2013){Astropy Collaboration},
  {Robitaille}, {Tollerud}, {Greenfield}, {Droettboom}, {Bray}, {Aldcroft},
  {Davis}, {Ginsburg}, {Price-Whelan}, {Kerzendorf}, {Conley}, {Crighton},
  {Barbary}, {Muna}, {Ferguson}, {Grollier}, {Parikh}, {Nair}, {Unther},
  {Deil}, {Woillez}, {Conseil}, {Kramer}, {Turner}, {Singer}, {Fox}, {Weaver},
  {Zabalza}, {Edwards}, {Azalee Bostroem}, {Burke}, {Casey}, {Crawford},
  {Dencheva}, {Ely}, {Jenness}, {Labrie}, {Lim}, {Pierfederici}, {Pontzen},
  {Ptak}, {Refsdal}, {Servillat}, \& {Streicher}}]{astropy:2013}
{Astropy Collaboration}, {Robitaille}, T.~P., {Tollerud}, E.~J., {et~al.} 2013,
  \aap, 558, A33, \dodoi{10.1051/0004-6361/201322068}

\bibitem[{{Astropy Collaboration} {et~al.}(2018){Astropy Collaboration},
  {Price-Whelan}, {Sip{\H{o}}cz}, {G{\"u}nther}, {Lim}, {Crawford}, {Conseil},
  {Shupe}, {Craig}, {Dencheva}, {Ginsburg}, {Vand erPlas}, {Bradley},
  {P{\'e}rez-Su{\'a}rez}, {de Val-Borro}, {Aldcroft}, {Cruz}, {Robitaille},
  {Tollerud}, {Ardelean}, {Babej}, {Bach}, {Bachetti}, {Bakanov}, {Bamford},
  {Barentsen}, {Barmby}, {Baumbach}, {Berry}, {Biscani}, {Boquien}, {Bostroem},
  {Bouma}, {Brammer}, {Bray}, {Breytenbach}, {Buddelmeijer}, {Burke},
  {Calderone}, {Cano Rodr{\'\i}guez}, {Cara}, {Cardoso}, {Cheedella}, {Copin},
  {Corrales}, {Crichton}, {D'Avella}, {Deil}, {Depagne}, {Dietrich}, {Donath},
  {Droettboom}, {Earl}, {Erben}, {Fabbro}, {Ferreira}, {Finethy}, {Fox},
  {Garrison}, {Gibbons}, {Goldstein}, {Gommers}, {Greco}, {Greenfield},
  {Groener}, {Grollier}, {Hagen}, {Hirst}, {Homeier}, {Horton}, {Hosseinzadeh},
  {Hu}, {Hunkeler}, {Ivezi{\'c}}, {Jain}, {Jenness}, {Kanarek}, {Kendrew},
  {Kern}, {Kerzendorf}, {Khvalko}, {King}, {Kirkby}, {Kulkarni}, {Kumar},
  {Lee}, {Lenz}, {Littlefair}, {Ma}, {Macleod}, {Mastropietro}, {McCully},
  {Montagnac}, {Morris}, {Mueller}, {Mumford}, {Muna}, {Murphy}, {Nelson},
  {Nguyen}, {Ninan}, {N{\"o}the}, {Ogaz}, {Oh}, {Parejko}, {Parley}, {Pascual},
  {Patil}, {Patil}, {Plunkett}, {Prochaska}, {Rastogi}, {Reddy Janga},
  {Sabater}, {Sakurikar}, {Seifert}, {Sherbert}, {Sherwood-Taylor}, {Shih},
  {Sick}, {Silbiger}, {Singanamalla}, {Singer}, {Sladen}, {Sooley},
  {Sornarajah}, {Streicher}, {Teuben}, {Thomas}, {Tremblay}, {Turner},
  {Terr{\'o}n}, {van Kerkwijk}, {de la Vega}, {Watkins}, {Weaver}, {Whitmore},
  {Woillez}, {Zabalza}, \& {Astropy Contributors}}]{astropy:2018}
{Astropy Collaboration}, {Price-Whelan}, A.~M., {Sip{\H{o}}cz}, B.~M., {et~al.}
  2018, \aj, 156, 123, \dodoi{10.3847/1538-3881/aabc4f}

\bibitem[{{Astropy Collaboration} {et~al.}(2022){Astropy Collaboration},
  {Price-Whelan}, {Lim}, {Earl}, {Starkman}, {Bradley}, {Shupe}, {Patil},
  {Corrales}, {Brasseur}, {N{"o}the}, {Donath}, {Tollerud}, {Morris},
  {Ginsburg}, {Vaher}, {Weaver}, {Tocknell}, {Jamieson}, {van Kerkwijk},
  {Robitaille}, {Merry}, {Bachetti}, {G{"u}nther}, {Aldcroft},
  {Alvarado-Montes}, {Archibald}, {B{'o}di}, {Bapat}, {Barentsen}, {Baz{'a}n},
  {Biswas}, {Boquien}, {Burke}, {Cara}, {Cara}, {Conroy}, {Conseil}, {Craig},
  {Cross}, {Cruz}, {D'Eugenio}, {Dencheva}, {Devillepoix}, {Dietrich},
  {Eigenbrot}, {Erben}, {Ferreira}, {Foreman-Mackey}, {Fox}, {Freij}, {Garg},
  {Geda}, {Glattly}, {Gondhalekar}, {Gordon}, {Grant}, {Greenfield}, {Groener},
  {Guest}, {Gurovich}, {Handberg}, {Hart}, {Hatfield-Dodds}, {Homeier},
  {Hosseinzadeh}, {Jenness}, {Jones}, {Joseph}, {Kalmbach}, {Karamehmetoglu},
  {Ka{l}uszy{'n}ski}, {Kelley}, {Kern}, {Kerzendorf}, {Koch}, {Kulumani},
  {Lee}, {Ly}, {Ma}, {MacBride}, {Maljaars}, {Muna}, {Murphy}, {Norman},
  {O'Steen}, {Oman}, {Pacifici}, {Pascual}, {Pascual-Granado}, {Patil},
  {Perren}, {Pickering}, {Rastogi}, {Roulston}, {Ryan}, {Rykoff}, {Sabater},
  {Sakurikar}, {Salgado}, {Sanghi}, {Saunders}, {Savchenko}, {Schwardt},
  {Seifert-Eckert}, {Shih}, {Jain}, {Shukla}, {Sick}, {Simpson},
  {Singanamalla}, {Singer}, {Singhal}, {Sinha}, {Sip{H{o}}cz}, {Spitler},
  {Stansby}, {Streicher}, {{{S}}umak}, {Swinbank}, {Taranu}, {Tewary},
  {Tremblay}, {Val-Borro}, {Van Kooten}, {Vasovi{'c}}, {Verma}, {de Miranda
  Cardoso}, {Williams}, {Wilson}, {Winkel}, {Wood-Vasey}, {Xue}, {Yoachim},
  {Zhang}, {Zonca}, \& {Astropy Project Contributors}}]{astropy:2022}
{Astropy Collaboration}, {Price-Whelan}, A.~M., {Lim}, P.~L., {et~al.} 2022,
  \apj, 935, 167, \dodoi{10.3847/1538-4357/ac7c74}

\bibitem[{{Ballot} {et~al.}(2021){Ballot}, {Roudier}, {Malherbe}, \&
  {Frank}}]{2021A&A...652A.103B}
{Ballot}, J., {Roudier}, T., {Malherbe}, J.~M., \& {Frank}, Z. 2021, \aap, 652,
  A103, \dodoi{10.1051/0004-6361/202039436}

\bibitem[{{Basu}(2016)}]{2016LRSP...13....2B}
{Basu}, S. 2016, Living Reviews in Solar Physics, 13, 2,
  \dodoi{10.1007/s41116-016-0003-4}

\bibitem[{{Bhatia} {et~al.}(2022){Bhatia}, {Cameron}, {Solanki}, {Peter},
  {Przybylski}, {Witzke}, \& {Shapiro}}]{Bhatia_SSD_I}
{Bhatia}, T.~S., {Cameron}, R.~H., {Solanki}, S.~K., {et~al.} 2022, \aap, 663,
  A166, \dodoi{10.1051/0004-6361/202243607}

\bibitem[{{Borrero} {et~al.}(2014){Borrero}, {Lites}, {Lagg}, {Rezaei}, \&
  {Rempel}}]{2014A&A...572A..54B}
{Borrero}, J.~M., {Lites}, B.~W., {Lagg}, A., {Rezaei}, R., \& {Rempel}, M.
  2014, \aap, 572, A54, \dodoi{10.1051/0004-6361/201424584}

\bibitem[{{Borrero} {et~al.}(2024){Borrero}, {Milic}, {Pastor Yabar},
  {Kaithakkal}, \& {de la Cruz Rodriguez}}]{BorMil2024_QS}
{Borrero}, J.~M., {Milic}, I., {Pastor Yabar}, A., {Kaithakkal}, A.~J., \& {de
  la Cruz Rodriguez}, J. 2024, arXiv e-prints, arXiv:2404.07067,
  \dodoi{10.48550/arXiv.2404.07067}

\bibitem[{{Borrero} {et~al.}(2019){Borrero}, {Pastor Yabar}, {Rempel}, \& {Ruiz
  Cobo}}]{borrero2019}
{Borrero}, J.~M., {Pastor Yabar}, A., {Rempel}, M., \& {Ruiz Cobo}, B. 2019,
  \aap, 632, A111, \dodoi{10.1051/0004-6361/201936367}

\bibitem[{{Brandt} \& {Solanki}(1990)}]{1990A&A...231..221B}
{Brandt}, P.~N., \& {Solanki}, S.~K. 1990, \aap, 231, 221

\bibitem[{Buehler {et~al.}(2013)Buehler, Lagg, \& Solanki}]{buehler_quiet_2013}
Buehler, D., Lagg, A., \& Solanki, S.~K. 2013, Astronomy \& Astrophysics, 555,
  A33, \dodoi{10.1051/0004-6361/201321152}

\bibitem[{Cavallini {et~al.}(1986)Cavallini, Ceppatelli, \&
  Righini}]{cavallini_long-term_1986}
Cavallini, F., Ceppatelli, G., \& Righini, A. 1986, Astronomy and Astrophysics,
  158, 275.
\newblock \url{https://ui.adsabs.harvard.edu/abs/1986A&A...158..275C}

\bibitem[{{Chatterjee} {et~al.}(2017){Chatterjee}, {Mandal}, \&
  {Banerjee}}]{2017ApJ...841...70C}
{Chatterjee}, S., {Mandal}, S., \& {Banerjee}, D. 2017, \apj, 841, 70,
  \dodoi{10.3847/1538-4357/aa709d}

\bibitem[{{Cheung} {et~al.}(2007){Cheung}, {Sch{\"u}ssler}, \&
  {Moreno-Insertis}}]{2007A&A...461.1163C}
{Cheung}, M.~C.~M., {Sch{\"u}ssler}, M., \& {Moreno-Insertis}, F. 2007, \aap,
  461, 1163, \dodoi{10.1051/0004-6361:20066390}

\bibitem[{{Criscuoli}(2013)}]{2013ApJ...778...27C}
{Criscuoli}, S. 2013, \apj, 778, 27, \dodoi{10.1088/0004-637X/778/1/27}

\bibitem[{{Criscuoli} \& {Uitenbroek}(2014)}]{2014ApJ...788..151C}
{Criscuoli}, S., \& {Uitenbroek}, H. 2014, \apj, 788, 151,
  \dodoi{10.1088/0004-637X/788/2/151}

\bibitem[{{Danilovic} {et~al.}(2008){Danilovic}, {Gandorfer}, {Lagg},
  {Sch{\"u}ssler}, {Solanki}, {V{\"o}gler}, {Katsukawa}, \&
  {Tsuneta}}]{Danilovic2008hinodevsmhd}
{Danilovic}, S., {Gandorfer}, A., {Lagg}, A., {et~al.} 2008, \aap, 484, L17,
  \dodoi{10.1051/0004-6361:200809857}

\bibitem[{{De Rosa} \& {Toomre}(2004)}]{2004ApJ...616.1242D}
{De Rosa}, M.~L., \& {Toomre}, J. 2004, \apj, 616, 1242, \dodoi{10.1086/424920}

\bibitem[{{Dravins} {et~al.}(1981){Dravins}, {Lindegren}, \&
  {Nordlund}}]{1981A&A....96..345D}
{Dravins}, D., {Lindegren}, L., \& {Nordlund}, A. 1981, \aap, 96, 345

\bibitem[{{Faurobert} {et~al.}(2016){Faurobert}, {Balasubramanian}, \&
  {Ricort}}]{2016A&A...595A..71F}
{Faurobert}, M., {Balasubramanian}, R., \& {Ricort}, G. 2016, \aap, 595, A71,
  \dodoi{10.1051/0004-6361/201527797}

\bibitem[{{Finsterle} {et~al.}(2013){Finsterle}, {Shapiro}, {Schmutz}, \&
  {Krivova}}]{2013EGUGA..1511672F}
{Finsterle}, W., {Shapiro}, A., {Schmutz}, W., \& {Krivova}, N. 2013, in EGU
  General Assembly Conference Abstracts, EGU General Assembly Conference
  Abstracts, EGU2013--11672

\bibitem[{{Fontenla} {et~al.}(1993){Fontenla}, {Avrett}, \& {Loeser}}]{falc}
{Fontenla}, J.~M., {Avrett}, E.~H., \& {Loeser}, R. 1993, \apj, 406, 319,
  \dodoi{10.1086/172443}

\bibitem[{Fouhey {et~al.}(2023)Fouhey, Higgins, Antiochos, Barnes, DeRosa,
  Hoeksema, Leka, Liu, Schuck, \& Gombosi}]{fouhey_large-scale_2023}
Fouhey, D.~F., Higgins, R. E.~L., Antiochos, S.~K., {et~al.} 2023, The
  Astrophysical Journal Supplement Series, 264, 49,
  \dodoi{10.3847/1538-4365/aca539}

\bibitem[{{Gingerich} {et~al.}(1971){Gingerich}, {Noyes}, {Kalkofen}, \&
  {Cuny}}]{1971SoPh...18..347G}
{Gingerich}, O., {Noyes}, R.~W., {Kalkofen}, W., \& {Cuny}, Y. 1971, \solphys,
  18, 347, \dodoi{10.1007/BF00149057}

\bibitem[{{Gray} \& {Livingston}(1997)}]{1997ApJ...474..802G}
{Gray}, D.~F., \& {Livingston}, W.~C. 1997, \apj, 474, 802,
  \dodoi{10.1086/303489}

\bibitem[{Grec {et~al.}(2010)Grec, Uitenbroek, Faurobert, \&
  Aime}]{grec_measuring_2010}
Grec, C., Uitenbroek, H., Faurobert, M., \& Aime, C. 2010, Astronomy and
  Astrophysics, 514, A91, \dodoi{10.1051/0004-6361/200811455}

\bibitem[{{Guenther} \& {Mattig}(1991)}]{1991A&A...243..244G}
{Guenther}, E., \& {Mattig}, W. 1991, \aap, 243, 244

\bibitem[{Harris {et~al.}(2020)Harris, Millman, van~der Walt, Gommers,
  Virtanen, Cournapeau, Wieser, Taylor, Berg, Smith, Kern, Picus, Hoyer, van
  Kerkwijk, Brett, Haldane, del R{\'{i}}o, Wiebe, Peterson,
  G{\'{e}}rard-Marchant, Sheppard, Reddy, Weckesser, Abbasi, Gohlke, \&
  Oliphant}]{harris2020array}
Harris, C.~R., Millman, K.~J., van~der Walt, S.~J., {et~al.} 2020, Nature, 585,
  357, \dodoi{10.1038/s41586-020-2649-2}

\bibitem[{{Hathaway}(2015)}]{2015LRSP...12....4H}
{Hathaway}, D.~H. 2015, Living Reviews in Solar Physics, 12, 4,
  \dodoi{10.1007/lrsp-2015-4}

\bibitem[{{Hinode Review Team} {et~al.}(2019){Hinode Review Team}, Al-Janabi,
  Antolin, Baker, Bellot~Rubio, Bradley, Brooks, Centeno, Culhane, Del~Zanna,
  Doschek, Fletcher, Hara, Harra, Hillier, Imada, Klimchuk, Mariska, Pereira,
  Reeves, Sakao, Sakurai, Shimizu, Shimojo, Shiota, Solanki, Sterling, Su,
  Suematsu, Tarbell, Tiwari, Toriumi, Ugarte-Urra, Warren, Watanabe, \&
  Young}]{hinode_review_team_achievements_2019}
{Hinode Review Team}, Al-Janabi, K., Antolin, P., {et~al.} 2019, Publications
  of the Astronomical Society of Japan, 71, R1, \dodoi{10.1093/pasj/psz084}

\bibitem[{Hunter(2007)}]{Hunter:2007}
Hunter, J.~D. 2007, Computing in Science \& Engineering, 9, 90,
  \dodoi{10.1109/MCSE.2007.55}

\bibitem[{{Janssen} \& {Cauzzi}(2006)}]{2006A&A...450..365J}
{Janssen}, K., \& {Cauzzi}, G. 2006, \aap, 450, 365,
  \dodoi{10.1051/0004-6361:20054310}

\bibitem[{{Jin} \& {Wang}(2015)}]{2015ApJ...807...70J}
{Jin}, C.~L., \& {Wang}, J.~X. 2015, ApJ, 807, 70,
  \dodoi{10.1088/0004-637X/807/1/70}

\bibitem[{{Kobel} {et~al.}(2012){Kobel}, {Solanki}, \&
  {Borrero}}]{2012A&A...542A..96K}
{Kobel}, P., {Solanki}, S.~K., \& {Borrero}, J.~M. 2012, \aap, 542, A96,
  \dodoi{10.1051/0004-6361/201118291}

\bibitem[{{Kopp}(2016)}]{2016JSWSC...6A..30K}
{Kopp}, G. 2016, Journal of Space Weather and Space Climate, 6, A30,
  \dodoi{10.1051/swsc/2016025}

\bibitem[{Korpi-Lagg {et~al.}(2022)Korpi-Lagg, Korpi-Lagg, Olspert, \&
  Truong}]{korpi-lagg_solar-cycle_2022}
Korpi-Lagg, M.~J., Korpi-Lagg, A., Olspert, N., \& Truong, H.-L. 2022,
  Solar-{Cycle} {Variation} of quiet-{Sun} {Magnetism} and {Surface} {Gravity}
  {Oscillation} {Mode}, \dodoi{10.1051/0004-6361/202243979}

\bibitem[{{Kuhn} {et~al.}(1988){Kuhn}, {Libbrecht}, \&
  {Dicke}}]{1988Sci...242..908K}
{Kuhn}, J.~R., {Libbrecht}, K.~G., \& {Dicke}, R.~H. 1988, Science, 242, 908,
  \dodoi{10.1126/science.242.4880.908}

\bibitem[{{Lefebvre} \& {Kosovichev}(2005)}]{2005ApJ...633L.149L}
{Lefebvre}, S., \& {Kosovichev}, A.~G. 2005, \apjl, 633, L149,
  \dodoi{10.1086/498305}

\bibitem[{{Leighton} {et~al.}(1962){Leighton}, {Noyes}, \&
  {Simon}}]{1962ApJ...135..474L}
{Leighton}, R.~B., {Noyes}, R.~W., \& {Simon}, G.~W. 1962, \apj, 135, 474,
  \dodoi{10.1086/147285}

\bibitem[{{Lites} {et~al.}(2014){Lites}, {Centeno}, \&
  {McIntosh}}]{2014PASJ...66S...4L}
{Lites}, B.~W., {Centeno}, R., \& {McIntosh}, S.~W. 2014, \pasj, 66, S4,
  \dodoi{10.1093/pasj/psu082}

\bibitem[{{Lites} {et~al.}(2013){Lites}, {Akin}, {Card}, {Cruz}, {Duncan},
  {Edwards}, {Elmore}, {Hoffmann}, {Katsukawa}, {Katz}, {Kubo}, {Ichimoto},
  {Shimizu}, {Shine}, {Streander}, {Suematsu}, {Tarbell}, {Title}, \&
  {Tsuneta}}]{2013SoPh..283..579L}
{Lites}, B.~W., {Akin}, D.~L., {Card}, G., {et~al.} 2013, \solphys, 283, 579,
  \dodoi{10.1007/s11207-012-0206-3}

\bibitem[{{Livingston}(1984)}]{1984ssdp.conf..330L}
{Livingston}, W. 1984, in Small-Scale Dynamical Processes in Quiet Stellar
  Atmospheres, ed. S.~L. {Keil}, 330

\bibitem[{{Livingston}(1987)}]{1987rfsm.conf...14L}
{Livingston}, W. 1987, in The Role of Fine-Scale Magnetic Fields on the
  Structure of the Solar Atmosphere, ed. E.~H. {Schr{\"o}ter},
  M.~{V{\'a}zquez}, \& A.~A. {Wyller}, 14

\bibitem[{{Livingston}(1982)}]{1982Natur.297..208L}
{Livingston}, W.~C. 1982, \nat, 297, 208, \dodoi{10.1038/297208a0}

\bibitem[{{Livingston}(1983)}]{1983IAUS..102..149L}
{Livingston}, W.~C. 1983, in IAU Symposium, Vol. 102, Solar and Stellar
  Magnetic Fields: Origins and Coronal Effects, ed. J.~O. {Stenflo}, 149--152

\bibitem[{MacQueen(1967)}]{macqueen1967some}
MacQueen, J.~B. 1967, in Proc. of the fifth Berkeley Symposium on Mathematical
  Statistics and Probability, ed. L.~M.~L. Cam \& J.~Neyman, Vol.~1 (University
  of California Press), 281--297

\bibitem[{{Macris} \& {Roesch}(1983)}]{1983CRASB.296..265M}
{Macris}, C.~J., \& {Roesch}, J. 1983, Academie des Sciences Paris Comptes
  Rendus Serie B Sciences Physiques, 296, 265

\bibitem[{{Magic} {et~al.}(2013){Magic}, {Collet}, {Asplund}, {Trampedach},
  {Hayek}, {Chiavassa}, {Stein}, \& {Nordlund}}]{2013A&A...557A..26M}
{Magic}, Z., {Collet}, R., {Asplund}, M., {et~al.} 2013, \aap, 557, A26,
  \dodoi{10.1051/0004-6361/201321274}

\bibitem[{{Mart{\'\i}nez Gonz{\'a}lez} {et~al.}(2006){Mart{\'\i}nez
  Gonz{\'a}lez}, {Collados}, \& {Ruiz Cobo}}]{2006A&A...456.1159M}
{Mart{\'\i}nez Gonz{\'a}lez}, M.~J., {Collados}, M., \& {Ruiz Cobo}, B. 2006,
  \aap, 456, 1159, \dodoi{10.1051/0004-6361:20065008}

\bibitem[{{Meunier} {et~al.}(2008){Meunier}, {Roudier}, \&
  {Rieutord}}]{2008A&A...488.1109M}
{Meunier}, N., {Roudier}, T., \& {Rieutord}, M. 2008, \aap, 488, 1109,
  \dodoi{10.1051/0004-6361:20078835}

\bibitem[{{Mili{\'c}} \& {van Noort}(2018)}]{snapi}
{Mili{\'c}}, I., \& {van Noort}, M. 2018, \aap, 617, A24,
  \dodoi{10.1051/0004-6361/201833382}

\bibitem[{{Mullan} {et~al.}(2007){Mullan}, {MacDonald}, \&
  {Townsend}}]{2007ApJ...670.1420M}
{Mullan}, D.~J., {MacDonald}, J., \& {Townsend}, R.~H.~D. 2007, \apj, 670,
  1420, \dodoi{10.1086/522559}

\bibitem[{{Muller}(1988)}]{1988AdSpR...8g.159M}
{Muller}, R. 1988, Advances in Space Research, 8, 159,
  \dodoi{10.1016/0273-1177(88)90186-X}

\bibitem[{{Muller} {et~al.}(2007){Muller}, {Hanslmeier}, \&
  {Salda{\~n}a-Mu{\~n}oz}}]{2007A&A...475..717M}
{Muller}, R., {Hanslmeier}, A., \& {Salda{\~n}a-Mu{\~n}oz}, M. 2007, \aap, 475,
  717, \dodoi{10.1051/0004-6361:20078387}

\bibitem[{{Muller} {et~al.}(2018){Muller}, {Hanslmeier}, {Utz}, \&
  {Ichimoto}}]{2018A&A...616A..87M}
{Muller}, R., {Hanslmeier}, A., {Utz}, D., \& {Ichimoto}, K. 2018, \aap, 616,
  A87, \dodoi{10.1051/0004-6361/201732085}

\bibitem[{{Nordlund} {et~al.}(2009){Nordlund}, {Stein}, \&
  {Asplund}}]{2009LRSP....6....2N}
{Nordlund}, {\r{A}}., {Stein}, R.~F., \& {Asplund}, M. 2009, Living Reviews in
  Solar Physics, 6, 2, \dodoi{10.12942/lrsp-2009-2}

\bibitem[{{Oba} {et~al.}(2017){Oba}, {Iida}, \&
  {Shimizu}}]{2017ApJ...836...40O}
{Oba}, T., {Iida}, Y., \& {Shimizu}, T. 2017, \apj, 836, 40,
  \dodoi{10.3847/1538-4357/836/1/40}

\bibitem[{Rempel(2018)}]{rempel_small-scale_2018}
Rempel, M. 2018, The Astrophysical Journal, 859, 161,
  \dodoi{10.3847/1538-4357/aabba0}

\bibitem[{{Rincon} \& {Rieutord}(2018)}]{2018LRSP...15....6R}
{Rincon}, F., \& {Rieutord}, M. 2018, Living Reviews in Solar Physics, 15, 6,
  \dodoi{10.1007/s41116-018-0013-5}

\bibitem[{{Roudier} {et~al.}(2017){Roudier}, {Malherbe}, \&
  {Mirouh}}]{2017A&A...598A..99R}
{Roudier}, T., {Malherbe}, J.~M., \& {Mirouh}, G.~M. 2017, \aap, 598, A99,
  \dodoi{10.1051/0004-6361/201629274}

\bibitem[{{Roudier} \& {Muller}(1986)}]{1986SoPh..107...11R}
{Roudier}, T., \& {Muller}, R. 1986, \solphys, 107, 11,
  \dodoi{10.1007/BF00155337}

\bibitem[{{Roudier} \& {Reardon}(1998)}]{1998ASPC..140..455R}
{Roudier}, T., \& {Reardon}, K. 1998, in Astronomical Society of the Pacific
  Conference Series, Vol. 140, Synoptic Solar Physics, ed. K.~S.
  {Balasubramaniam}, J.~{Harvey}, \& D.~{Rabin}, 455

\bibitem[{Ruiz~Cobo \& del Toro~Iniesta(1992)}]{ruiz_cobo_inversion_1992}
Ruiz~Cobo, B., \& del Toro~Iniesta, J.~C. 1992, The Astrophysical Journal, 398,
  375, \dodoi{10.1086/171862}

\bibitem[{{Sainz Dalda} {et~al.}(2024){Sainz Dalda}, {Agrawal}, {De Pontieu},
  \& {Go{\v{s}}i{\'c}}}]{2024ApJS..271...24S}
{Sainz Dalda}, A., {Agrawal}, A., {De Pontieu}, B., \& {Go{\v{s}}i{\'c}}, M.
  2024, \apjs, 271, 24, \dodoi{10.3847/1538-4365/ad1e55}

\bibitem[{{Scherrer} {et~al.}(2012){Scherrer}, {Schou}, {Bush}, {Kosovichev},
  {Bogart}, {Hoeksema}, {Liu}, {Duvall}, {Zhao}, {Title}, {Schrijver},
  {Tarbell}, \& {Tomczyk}}]{2012SoPh..275..207S}
{Scherrer}, P.~H., {Schou}, J., {Bush}, R.~I., {et~al.} 2012, \solphys, 275,
  207, \dodoi{10.1007/s11207-011-9834-2}

\bibitem[{Smitha {et~al.}(2020)Smitha, Holzreuter, van Noort, \&
  Solanki}]{Smitha_2020}
Smitha, H.~N., Holzreuter, R., van Noort, M., \& Solanki, S.~K. 2020, Astronomy
  \& Astrophysics, 633, A157, \dodoi{10.1051/0004-6361/201937041}

\bibitem[{{Spada} {et~al.}(2018){Spada}, {Arlt}, {K{\"u}ker}, \&
  {Sofia}}]{2018AN....339..545S}
{Spada}, F., {Arlt}, R., {K{\"u}ker}, M., \& {Sofia}, S. 2018, Astronomische
  Nachrichten, 339, 545, \dodoi{10.1002/asna.201813521}

\bibitem[{{Trampedach} {et~al.}(2013){Trampedach}, {Asplund}, {Collet},
  {Nordlund}, \& {Stein}}]{2013ApJ...769...18T}
{Trampedach}, R., {Asplund}, M., {Collet}, R., {Nordlund}, {\r{A}}., \&
  {Stein}, R.~F. 2013, \apj, 769, 18, \dodoi{10.1088/0004-637X/769/1/18}

\bibitem[{{Trelles Arjona} {et~al.}(2023){Trelles Arjona}, {Mart{\'\i}nez
  Gonz{\'a}lez}, \& {Ruiz Cobo}}]{Trelles2023QS}
{Trelles Arjona}, J.~C., {Mart{\'\i}nez Gonz{\'a}lez}, M.~J., \& {Ruiz Cobo},
  B. 2023, \apj, 944, 95, \dodoi{10.3847/1538-4357/acb64d}

\bibitem[{{Tsuneta} {et~al.}(2008){Tsuneta}, {Ichimoto}, {Katsukawa}, {Nagata},
  {Otsubo}, {Shimizu}, {Suematsu}, {Nakagiri}, {Noguchi}, {Tarbell}, {Title},
  {Shine}, {Rosenberg}, {Hoffmann}, {Jurcevich}, {Kushner}, {Levay}, {Lites},
  {Elmore}, {Matsushita}, {Kawaguchi}, {Saito}, {Mikami}, {Hill}, \&
  {Owens}}]{2008SoPh..249..167T}
{Tsuneta}, S., {Ichimoto}, K., {Katsukawa}, Y., {et~al.} 2008, \solphys, 249,
  167, \dodoi{10.1007/s11207-008-9174-z}

\bibitem[{Uitenbroek \& Criscuoli(2011)}]{uitenbroek_why_2011}
Uitenbroek, H., \& Criscuoli, S. 2011, The Astrophysical Journal, 736, 69,
  \dodoi{10.1088/0004-637X/736/1/69}

\bibitem[{{Vesa} {et~al.}(2023){Vesa}, {Jackiewicz}, \&
  {Reardon}}]{2023ApJ...952...58V}
{Vesa}, O., {Jackiewicz}, J., \& {Reardon}, K. 2023, \apj, 952, 58,
  \dodoi{10.3847/1538-4357/acd930}

\bibitem[{Virtanen {et~al.}(2020)Virtanen, Gommers, Oliphant, Haberland, Reddy,
  Cournapeau, Burovski, Peterson, Weckesser, Bright, {van der Walt}, Brett,
  Wilson, Millman, Mayorov, Nelson, Jones, Kern, Larson, Carey, Polat, Feng,
  Moore, {VanderPlas}, Laxalde, Perktold, Cimrman, Henriksen, Quintero, Harris,
  Archibald, Ribeiro, Pedregosa, {van Mulbregt}, \& {SciPy 1.0
  Contributors}}]{2020SciPy-NMeth}
Virtanen, P., Gommers, R., Oliphant, T.~E., {et~al.} 2020, Nature Methods, 17,
  261, \dodoi{10.1038/s41592-019-0686-2}

\bibitem[{{Watson} \& {Basu}(2020)}]{2020ApJ...903L..29W}
{Watson}, C.~B., \& {Basu}, S. 2020, \apjl, 903, L29,
  \dodoi{10.3847/2041-8213/abc348}

\bibitem[{{Yeo} {et~al.}(2017){Yeo}, {Solanki}, {Norris}, {Beeck}, {Unruh}, \&
  {Krivova}}]{2017PhRvL.119i1102Y}
{Yeo}, K.~L., {Solanki}, S.~K., {Norris}, C.~M., {et~al.} 2017, \prl, 119,
  9.1102, \dodoi{10.1103/PhysRevLett.119.091102}

\end{thebibliography}
\bibliographystyle{aasjournal}

%\appendix
%\section{ data and results}\label{sec:appendix}

\begin{table} 
\caption{Hinode SOT/SP rasters used in this paper. For all dates, the FoV analyzed was at disk center, and 51.8'' $\times$ 24.5'' wide.}
\label{table:datasets}
\centering
\begin{longtable}{|p{1.5cm}||  p{2.5cm}| p{2cm}|}
 \hline
%\multicolumn{3}{|c|}{Hinode SOT/SP rasters used in this paper.} \\ 
%\hline
Dataset Number & Date & Time (UT) \\ \hline
 1 & 2008/01/30 & 22:14:05\\ \hline
 2 & 2008/12/17 & 04:59:05\\ \hline
 3 & 2009/01/23 & 10:44:05\\ \hline
 4 & 2009/02/17 & 22:34:05\\ \hline
 5 & 2009/04/11 & 18:05:06\\ \hline
 6 & 2009/07/28 & 15:29:06\\ \hline
 7 & 2009/12/02 & 16:04:06\\ \hline
 8 & 2010/05/31 & 02:00:26\\ \hline
 9 & 2010/10/07 & 15:08:06\\ \hline
 10 & 2010/12/02 & 16:33:05\\ \hline
 11 & 2011/02/02 & 13:47:05\\ \hline
 12 & 2011/06/30 & 20:38:35\\ \hline
 13 & 2011/11/30 & 21:38:35\\ \hline
 14 & 2012/03/31 & 01:23:05\\ \hline
 15 & 2012/07/26 & 18:57:06\\ \hline
 16 & 2013/01/22 & 14:12:06\\ \hline
 17 & 2013/03/26 & 14:17:36\\ \hline
 18 & 2013/07/03 & 07:35:04\\ \hline
 19 & 2013/09/18 & 15:57:20\\ \hline
 20 & 2013/12/25 & 14:55:35\\ \hline
 21 & 2014/08/19 & 21:37:35\\ \hline
 22 & 2015/08/14 & 23:06:04\\ \hline
 23 & 2015/09/15 & 21:19:05\\ \hline
 24 & 2015/12/15 & 17:47:36\\ \hline
 25 & 2016/02/17 & 13:22:05\\ \hline
 26 & 2016/07/23 & 17:49:05\\ \hline
 27 & 2016/11/15 & 15:49:05\\ \hline
 28 & 2017/02/19 & 02:58:05\\ \hline
 29 & 2017/05/18 & 17:43:06\\ \hline
 30 & 2017/11/18 & 16:35:36\\ \hline
 31 & 2017/12/14 & 13:43:57\\ \hline
 32 & 2018/04/13 & 12:25:04 \\ \hline
 33 & 2018/08/19 & 13:53:35\\ \hline
 34 & 2018/11/15 & 17:22:06\\ \hline
 35 & 2019/06/13 & 19:09:06\\ \hline
 36 & 2019/12/13 & 16:53:05\\ \hline
 37 & 2021/01/15 & 03:47:35\\ \hline
 38 & 2021/02/18 & 00:19:34\\ \hline
 39 & 2021/03/18 & 23:43:05\\ \hline
 40 & 2021/04/15 & 15:42:36\\ \hline
 41 & 2021/06/17 & 16:59:36\\ \hline
 42 & 2021/08/13 & 15:03:05\\ \hline
 43 & 2021/12/16 & 18:58:06\\ \hline
 44 & 2022/04/14 & 15:40:06\\ \hline
 45 & 2022/04/14 & 16:01:06\\ \hline
 46 & 2022/06/09 & 17:50:35\\ \hline
 47 & 2022/07/21 & 15:53:36\\ \hline
 48 & 2022/10/20 & 14:49:34\\ \hline
 49 & 2022/12/28 & 11:42:05\\ \hline
 \hline
\end{longtable}

\end{table}

\end{document}